\documentclass[11pt]{article}
\pdfoutput=1
\usepackage{amsmath}
\usepackage{amssymb}
\usepackage{graphicx,bbm,mathrsfs}
\usepackage{nicefrac}
\usepackage{bbm}
\usepackage{geometry}
\usepackage{dcolumn}   
\usepackage{bm}        
\usepackage{graphicx,mathrsfs}
\usepackage{nicefrac}
\usepackage{multirow}
\usepackage{color}
\usepackage{relsize}
\usepackage{adjustbox}

\newcommand{\be}{\begin{equation}}
\newcommand{\ee}{\end{equation}}
\newcommand{\bea}{\begin{eqnarray}}
\newcommand{\eea}{\end{eqnarray}}

\usepackage{wasysym}

\begin{document}

\vspace*{1.2cm}

\begin{center}

\thispagestyle{empty}
{\LARGE
Opacity and Effective Field Theory in \\ Anti-de Sitter Backgrounds
 }\\[10mm]

\renewcommand{\thefootnote}{\fnsymbol{footnote}}

{\large  Sylvain~Fichet
\footnote{sylvain.fichet@gmail.com } }\\[10mm]

\end{center} 
\noindent
\quad\quad\quad\textit{Walter Burke Institute for Theoretical Physics, California Institute of Technology,}

\noindent \quad\quad\quad \textit{Pasadena, CA
91125, California, USA} \\

\noindent
\quad\quad\quad \textit{ICTP South American Institute for Fundamental Research  \& IFT-UNESP,}

\noindent \quad\quad\quad \textit{R. Dr. Bento Teobaldo Ferraz 271, S\~ao Paulo, Brazil
}

\addtocounter{footnote}{-1}

\vspace*{12mm}

\begin{center}
{  \bf  Abstract }
\end{center}

We consider  quantum field theory in a five-dimensional anti-de Sitter background, possibly truncated by 4d branes.
 In the Euclidian version of this space, it is known that propagators 
 exponentially decay when they go far enough towards the Poincar\'e horizon, \textit{i.e.} deep enough into the IR region of the AdS bulk.
   In this note we show that  an analogous property exists in Lorentzian AdS. 
  The exponential suppression is found to occur in the presence of bulk interactions  dressing the AdS propagators.
    We calculate one-loop gravitational dressing  and find that the suppression effect comes from the scalar component of the 5d graviton. 
We then argue that, at least at strong coupling,  this exponential decay censors the region of spacetime where the 5d effective field theory would become invalid.  
As an application we estimate the rate for the cascade decay  of bulk fields---which are known to produce soft spherical events with high multiplicity---and find  this rate to be exponentially suppressed.

\section{Introduction \label{se:intro}}

Quantum field theory in anti de-Sitter spacetime   has been a source of deep formal insights  as well as elegant phenomenological developments for at least two decades. 
Formally, investigations of the AdS/CFT conjecture and its holographic realization have brought knowledge about the theories on both sides of the correspondence \cite{Maldacena:1997re,Witten:1998qj,Aharony:1999ti}. Phenomenologically, AdS space and its truncated or deformed versions
have offered a rich playground for physics beyond the Standard Model, providing for instance a simple explanation to the electroweak hierarchy problem \cite{Randall:1999ee}.


In five dimensions, the field theory living in the AdS background should be seen as an effective field theory (EFT) valid up to some cutoff on short 5d distances $|\Delta X^M| \sim 1/\Lambda $,  
 beyond which the whole 5d model is superseded by a theory of quantum gravity. 
When described in the Poincar\'e patch with conformal coordinates $X^M=(x^\mu,z)$, a striking difference between  AdS space and flat space is that the validity limit of the EFT is also reached for small enough 4d distance $|x^\mu|$  in such a way that the validity cutoff of the EFT depends on the  position in the bulk  of AdS. Namely, the validity cutoff of the theory scales along the fifth dimension  as $1/z$. 
This essential fact as long been known (see \textit{e.g.} \cite{Randall:1999ee,Randall:2001gb}),  and it is perhaps in \cite{Goldberger:2002cz} that  it has been made clear that the position-dependent cutoff simply results from the presence of higher dimensional operators in the effective 5d theory.

Another feature inherent to curved space, known in the Euclidian version of AdS, is that propagators becomes exponentially suppressed when one of the endpoints in the fifth dimension goes too far in the IR. 
In position-momentum space, if a propagator carries an Euclidian absolute four-momentum $p$, the exponential suppression occurs beyond $p\sim 1/z $  \cite{Goldberger:2002cz}.
Interestingly, this region matches roughly the region in which the 5d EFT would break down. Therefore in Euclidian AdS it turns out that  propagators refuse to go in the region where the EFT is invalid: in a sense the theory censors itself. 

However Euclidian AdS is of limited interest since it does not contain particles. 
A more interesting question is whether the propagators have a similar behaviour in Lorentzian AdS. 
Since the position-dependent cutoff is also present in Lorentzian AdS, 
one may wonder if a similar mechanism of  ``censorship'' of the IR region may happen, which would require a decay of the propagator in the IR  as in the Euclidian case.

This question is the main focus of the present note: we  study how and to what extent the propagators decay in Lorentzian AdS.  This is a  feature of relevance  both for the sake of thoroughly understanding effective field theory in AdS   and in view of subsequent developments for physics beyond the Standard Model such as a ``warped dark sector'' \cite{mypaper}. 
The steps taken in the paper are as follows. In Sec.~\ref{se:setup} we lay down the AdS setup  needed for our analysis.  The behaviour of the scalar propagator in the different regions of position-momentum space is studied in Sec.~\ref{se:lims}.  We  then turn our attention to the effect of the imaginary part of the 1PI insertions dressing the propagator. 
After introducing a  Kaluza-Klein sum trick in Sec.~\ref{se:sums}, we  investigate the simple case of dressing by a cubic scalar interaction in Sec.~\ref{se:scalar}. The dressing of the propagator by 5d gravity is then evaluated in Sec.~\ref{se:grav}.  Comments about  censorship of the EFT are made in Sec.~\ref{se:EFT}.  As a corollary to our study, and taking profit of the formalism,  we  study the rate of  cascade decay of bulk fields in Sec.~\ref{se:decay}.  Sec.~\ref{se:con} contains our conclusions and details on the graviton loop calculation are given in the Appendix.

\section{A slice of AdS \label{se:setup}}

The full action used in this paper describes a scalar field and gravity in a slice of AdS$_5$. The action takes the form 
\be \label{eq:5d_action}
S_{\rm AdS}= \int d^5X^M \sqrt{|g|}\left[ M^3_* \left({\cal R} - \Lambda\right)+{\cal L}_\Phi \right]  \,,\ee
where spacetime is taken to be AdS$_5$ with cosmological constant $\Lambda=-12 k^2$, $k$ being the AdS curvature. The $M_*$ parameter sets the strength of 5d gravity.\footnote{With this convention we  have the matching $M_*^3\approx kM^2_{\rm Pl}$ in the 4d low-energy theory. }  $\cal R$ is the Ricci scalar  and ${\cal L}_{\Phi}$ describes a 5d scalar field.   
The graviton Lagrangian is expanded in Sec.~\ref{se:grav}. In this section  our focus is instead on the scalar field. The metric of the AdS background is denoted $\gamma_{MN}$, such that $g_{MN}=\gamma_{MN}+\ldots$ where the ellipse denotes the metric fluctuations.

 We use the Poincar\'e patch with conformal coordinates 
\be
ds^2=\gamma_{MN}dX^MdX^N=(kz)^{-2}(\eta_{\mu\nu}x^\mu x^\nu-dz^2)
\ee
where $\eta_{\mu\nu}$ is Minkowski metric with $(+,-,-,-)$ signature. The values $z=0$, $z\rightarrow \infty$ correspond respectively to the AdS boundary and to the Poincar\'e horizon.  The fifth dimension is assumed to be compact with $z\in[z_0,z_1]$, where $z_0\equiv 1/k$,   $z_1\equiv 1/\mu$ are respectively referred to as UV and IR branes.

 The Lagrangian for the scalar reads
\be
{\cal L}_\Phi = \frac{1}{2}\nabla_M \Phi \nabla^M \Phi -\frac{1}{2} m_\Phi^2 \Phi^2 + {\cal L}_{\rm int}+{\cal L}_{\cal B}\,,
\ee
where $m^2_\Phi$ is the scalar bulk mass, ${\cal L}_{\rm int}$ describes bulk matter interactions, and ${\cal L}_{\cal B}$ describes 4d brane-localized Lagrangians.  
The scalar bulk mass has to satisfy the Breitenlohner-Freedman bound $m_\Phi^2 \geq -4k^2$ to prevent tachyonic instabilities in AdS$_5$ \cite{Breitenlohner:1982bm}. 

Extremizing the action in $\Phi$ gives the classical 5d equation of motion
\be
{\cal D}\Phi = 0 \label{eq:5dEOMhom}
\ee
with the operator
\be
{\cal D} = \frac{1}{\sqrt{\gamma}} \partial_M \left( \gamma^{MN} \sqrt{\gamma} \partial_N \right) + m^2_\Phi\,,
\label{eq:5dop}
\ee
and two boundary conditions
\be
{\cal B}^{\rm UV} \Phi |_{z_0}  =0\,,\quad 
{\cal B}^{\rm IR} \Phi |_{z_1}  =0\,, \label{eq:BCsgen}
\ee
where the ${\cal B}^{\rm UV,\,IR}$ operators depend on the brane localized Lagrangians. 
The Feynman propagator $\Delta(X,X') $ of the free scalar in the AdS background is the Green function satisfying the equation of motion
\be
{\cal D}_X \Delta(X,X') = - i \frac{ \delta^{(5)}(X-X')} {\sqrt{\gamma}} \label{eq:5dEOMgreen}
\ee
as well as the boundary conditions Eqs.~\eqref{eq:BCsgen}. 

From now on we work in momentum space along the 4d Minkowski directions  such that the coordinates are $(p^\mu,z)$. The Greek indexes are understood to be contracted using the $\eta_{\mu\nu}$ metric.  We have $\partial_\mu \partial^\mu \Phi=-p_\mu p^\mu \Phi$. 
 On introduces $p=\sqrt{p_\mu p^\mu}$, which is real (imaginary) for timelike (spacelike) 4-momentum $p^\mu$. The solutions to the equation of motion Eq.~\eqref{eq:5dEOMhom} in the 5d position-momentum space are given by
\be
z^2 J_\alpha(p z)\,,\quad z^2 Y_\alpha(p z)\,, \label{eq:solsAdS}
\ee
where the parameter $\alpha \in \mathbf{R}$ is related to the scalar bulk mass by $\alpha^2 = \frac{m_\Phi^2}{k^2}+4 $. The boundary operators applied to the solutions Eq.~\eqref{eq:solsAdS} are denoted 
\footnote{The overall factors of $\tilde J^{\rm UV}_{\alpha}$, $\tilde J^{\rm IR}_{\alpha}$ are 
arbitrary, since the the boundary operators are defined up to a constant. These overall constants cancel inside the expression of the propagator.
}
\be
{\cal B}^{\rm UV,IR} [z^2 J_\alpha(pz)]_{z_0,z_1}  \equiv \tilde J^{\rm UV, IR}_\alpha\,, \quad 
{\cal B}^{\rm UV,IR} [z^2 Y_\alpha(pz)]_{z_0,z_1}  \equiv \tilde Y^{\rm UV, IR}_\alpha\,.
\ee
With these definitions the Feynman propagator is 
\begin{align}
\left\langle  \Phi(p,z) \Phi(-p,z') \right\rangle & \equiv \Delta_p(z,z')=
\\
i\frac{\pi k^3 (zz')^2}{2 }   & 
\frac{
\left[\tilde Y^{\rm UV}_{\alpha}J_{\alpha}\left(pz_<\right)
- \tilde J^{\rm UV}_{\alpha} Y_{\alpha}\left(pz_<
\right)\right]\left[
\tilde Y^{\rm IR}_{\alpha}J_{\alpha}\left(pz_>\right)
- \tilde J^{\rm IR}_{\alpha} Y_{\alpha}\left(pz_>
\right)
\right]}
{\tilde J^{\rm UV}_{\alpha} \tilde  Y^{\rm IR}_{\alpha}
- \tilde  Y^{\rm UV}_{\alpha} \tilde  J^{\rm IR}_{\alpha}}\,
\label{eq:propa_gen}
\end{align}
where $z_<={\rm min} (z,z')$, $z_>={\rm max} (z,z')$.

Finally  let us specify the brane terms. For our purposes it is enough to consider 
\be
{\cal L}_{\cal B} =   \sqrt{\bar \gamma} \,\frac{1}{2}\Phi^2\left(
\delta(z-z_0)\,M_{\rm UV}  - \delta(z-z_1) \,M_{\rm IR} 
\right) \, \label{eq:5d_bound_action}
\ee
where $\bar \gamma_{\mu\nu}$ is the induced metric on the branes, giving $\sqrt{\bar \gamma}=(kz)^{-4}$.
It is convenient to express the brane masses as
\be
M_{\rm UV}=  (\alpha-2)k - b_{\rm UV}k \,, \quad M_{\rm IR}=  (\alpha-2)k +  b_{\rm IR}k \,. \ee
The boundary conditions Eq.~\eqref{eq:BCsgen} then give
\be
\tilde J^{\rm UV}_{\alpha} =  \frac{p}{k} \, J_{\alpha-1}\left(\frac{p}{k}\right) - b_{\rm UV} \,J_\alpha\left(\frac{p}{k}\right)\,,\quad \tilde J^{\rm IR}_{\alpha}  =  \frac{p}{\mu} \, J_{\alpha-1}\left(\frac{p}{\mu}\right) + b_{\rm IR} \,J_\alpha\left(\frac{p}{\mu}\right)\,,
\ee
and similarly for the $Y_\alpha$ functions.

In these conventions, a massless mode is present in the spectrum when $b_i=0$ or $b_i=2\alpha$. Note  this becomes manifest when  $b_{\rm UV}=b_{\rm IR}=0$, $\alpha=2$ or $b_{\rm UV}=-b_{\rm IR}=-2\alpha$, $\alpha=-2$, since in both cases one has $m_\Phi=0$, $M_i=0$ so that all mass terms vanish in the 5d Lagrangian. The sign of the $b_i$'s is chosen such that  positive $b_i$ means a positive mass contribution to the 4d modes.

\section{Limits of the propagator \label{se:lims}}

Here we study the behaviour of the propagator $\Delta_p\left(z,z'\right)$ in the various regions of  position-momentum space. 
We can notice that the change $(b_{\rm UV}\rightarrow b_{\rm UV}+2\alpha,\, b_{\rm IR}\rightarrow b_{\rm IR}-2\alpha$) reverses the sign of $\alpha$ in the boundary conditions. A subsequent change $\alpha\rightarrow -\alpha$ gives back the original boundary condition, and we can check that the full propagator remains unchanged under these operations. Hence there is a symmetry in the solutions and for our purpose it is convenient to focus on $\alpha\geq 0$. The $b_{\rm UV}, b_{\rm IR}$ parameters
do not play an important role for this work and will remain unspecified.

For $p$ real (\textit{i.e.} $p^\mu$ timelike) and larger than $\mu$, the propagator has an infinite series of Kaluza Klein (KK) poles with $O(\mu)$ spacing,  whose masses are given by the equation \be
{\tilde J^{\rm UV}_{\alpha} \tilde  Y^{\rm IR}_{\alpha}
- \tilde  Y^{\rm UV}_{\alpha} \tilde  J^{\rm IR}_{\alpha}}=0\,.
\ee 
The propagator can be written as an infinite sum over the KK modes but this representation is not of immediate use, our focus here is rather on the closed form of the propagator.

We first consider the low-momentum region $|p|\lesssim\mu$ (\textit{Region 1}), where we have
\be
\Delta_p\left(z,z'\right)\approx i
\frac{k^3 (zz')^{2-\alpha }  \left(2 \alpha -b_{\rm IR} +b_{\rm IR} (T z_>)^{2 \alpha }\right) \left(2 \alpha+b_{\rm UV} -b_{\rm UV}
   (k z_<)^{2 \alpha }\right)}{2 \alpha  \left(b_{\rm UV} (2 \alpha -b_{\rm IR}) k^{2 \alpha }+b_{\rm IR} (2 \alpha +b_{\rm UV}) T^{2
   \alpha }\right)}  
   \,.
   \label{eq:propgen}
\ee
We can see that this  term is constant with respect to $p$ and represents a contact operator induced by the  heavy KK modes. The expression diverges at $b_{\rm UV,IR} = 0$ or $b_{ \rm UV}= - b_{ \rm IR} = -2\alpha$, signaling that a massless mode is also present in these particular cases.

Increasing the momentum, we have the region $\mu\lesssim |p|\lesssim\frac{1}{z_>}$ (\textit{Region 2}), in which  the KK poles along the real axis appear. Let us study the form of the propagator \textit{away} from the poles,
\begin{align}
& \Delta_p\left(z,z'\right)\approx i
\frac{(kz)^2(kz')^2}{2\alpha \, b_{\rm UV}\,k  } 
\frac{1}{(z_>k)^\alpha}
\left( 
\frac{b_{\rm UV}+2\alpha }{(z_<k)^\alpha}
-b_{\rm UV} (z_<k)^\alpha
 \right) \nonumber \\
& \quad - i\frac{\Gamma(-\alpha)\,(kz)^2(kz')^2}{\Gamma(\alpha+1)\,2b^2_{\rm UV}\,k}
\left(\frac{b_{\rm UV}+2\alpha }{(z_>k)^\alpha}
-b_{\rm UV} (z_>k)^\alpha
 \right) 
\left(\frac{b_{\rm UV}+2\alpha }{(z_<k)^\alpha}
-b_{\rm UV} (z_<k)^\alpha
 \right) 
\left( \frac{-p^2}{4k^2}\right)^\alpha
 \,.
\label{eq:propaII+}
\end{align}
It turns out that, away from the poles, the expression becomes independent of $\mu$ and $b_{\rm IR}$.~\footnote{
To be slightly more precise, along the real axis and away from the poles the second term is proportional 
to $S_\alpha=\frac{\sin\left(\frac{p}{\mu}-\frac{\pi}{4}(1-2\alpha)\right)}{\sin\left(\frac{p}{\mu}-\frac{\pi}{4}(1+2\alpha)\right)}$, which has periodic zeros. But whenever $p$ gets an imaginary part larger than $\sim \pi \alpha \mu $, one gets $S_\alpha \approx (-1)^{\alpha} $, which has been used in Eq.~\eqref{eq:propaII+}.  }
This signals the fact that the existence of the IR brane is  irrelevant away from the poles. The IR boundary condition, unlike the UV one, has no impact on the propagator in this regime, and  the IR brane can be sent to infinity ($1/\mu \rightarrow \infty$) without affecting Eq.~\eqref{eq:propaII+}.

The first term in Eq.~\eqref{eq:propaII+} has almost the same form as in the low energy limit except that $b_{\rm IR}$ is set to zero. From this we conclude that this term somehow encodes the effect from the heavy modes.
In contrast the second term is non-analytic in $p$ and one may thus suspect that it encodes the collective effect of  light modes.  
This is easily verified by for instance evaluating the non-relativistic potential induced by the $t$-channel exchange of the $\Phi$ continuum between static sources placed at points $z$, $z'$ in the bulk, and with three-dimensional separation  $r=|{\mathbf r}|$. 
When evaluating the spatial potential, the first term of Eq.~\eqref{eq:propaII+} is analytic and contribute as a delta function $\delta({r})$, while the second, non-analytic term induces a non-local contribution
\be
V({ r}) \propto \frac{1}{r} \frac{1}{(kr)^{2+2\alpha}}\,.
\ee
Since a potential with finite range must  result from the exchange of light degrees of freedom, we conclude that the non-analytic term of Eq.~\eqref{eq:propaII+} captures the collective contribution of light KK modes.~\footnote{
One may  notice that in the context of AdS/CFT it is again the non-analytic term which plays the key role. Namely,  in the UV-to-UV brane propagator $\Delta_p\left(z_0,z_0\right)$, this term  encodes the CFT operator probed by the $\Phi(z_0)$ source. In contrast, the first term of 
Eq.~\eqref{eq:propaII+} only gives a mass to the source.  }


Increasing again the 4-momentum we then enter the region $\frac{1}{z_>}\lesssim|p|\lesssim\frac{1}{z_<}$ (\textit{Region 3})
where the propagator takes the form
\be
\Delta_q\left(z,z'\right)=-i
\sqrt{\pi}(kz_<)^2 (k z_>)^{3/2}\frac{b_{\rm UV}(kz_<)^{\alpha}-(b_{\rm UV}+2\alpha)(kz_<)^{-\alpha} }{2b_{\rm UV}\,k\,\Gamma(\alpha+1)}\left(\frac{p}{2k}\right)^{\alpha-1/2} 
\frac{\cos\left(
\frac{p}{\mu}-pz_> \right)
}{\cos\left(
\frac{p}{\mu}+\frac{\pi}{4}(1-2\alpha) \right)}\,.
\ee      
This expression is non-analytic, but more importantly we can see that whenever $p$ has an imaginary part the whole expression is exponentially suppressed as
\be
\Delta_p\left(z,z'\right) \propto e^{- |{\rm  Im} \, p| z_>}\,. \label{eq: Delta_exp}
\ee 
 For spacelike 4-momentum, where $p$ is purely imaginary,  the exponential suppression Eq.~\eqref{eq: Delta_exp} occurs just like in Euclidian AdS  \cite{Gherghetta:2003he}. Instead, for timelike momentum, the propagator is merely oscillating  and no exponential suppression occurs. However, we may suspect that the situation changes in the  presence of interactions, which  dress the propagator with 1PI insertions and induce an imaginary component. A suppression of the kind of Eq.~\eqref{eq: Delta_exp} is nevertheless not guaranteed.
Hence a full calculation is required for knowing the behaviour of the dressed propagator, which will be carried out in Secs.~\ref{se:scalar},~\ref{se:grav}. This behaviour has no equivalent in flat space.

Finally, at even higher momentum, $\frac{1}{z_<}\lesssim|p|$ (\textit{Region 4}), we have
 \be
\Delta_p\left(z,z'\right)= i \frac{(kz)^{3/2}(kz')^{3/2}}{ p  }
\frac{ \cos \left(p \left(\frac{1}{T}-z_>\right)\right) \cos
   \left(p z_< - \frac{\pi}{4} (1+2   \alpha  )\right)}{
\sin \left(\frac{ p}{T}-\frac{\pi}{4} \left(1+ 2   \alpha  \right)\right)   
   } \,.
 \ee
As well known \cite{Randall:2001gb}, this propagator resembles very much the flat space propagator in a 5d compact dimension of radius $1/T$, up to the constant phases and to the non-trivial $(zz)^{3/2}$ factor. 
  Away from the real axis, this propagator is exponentially suppressed as
 \be
\Delta_p\left(z,z'\right) \propto e^{- |{\rm  Im} \, p| |z-z'|}\,. \label{eq: Delta_exp_4}
\ee

\section{AdS mode sums \label{se:sums}}

\begin{figure}
\center
\includegraphics[width=12 cm,trim={1cm 7cm 3cm 6cm},clip]{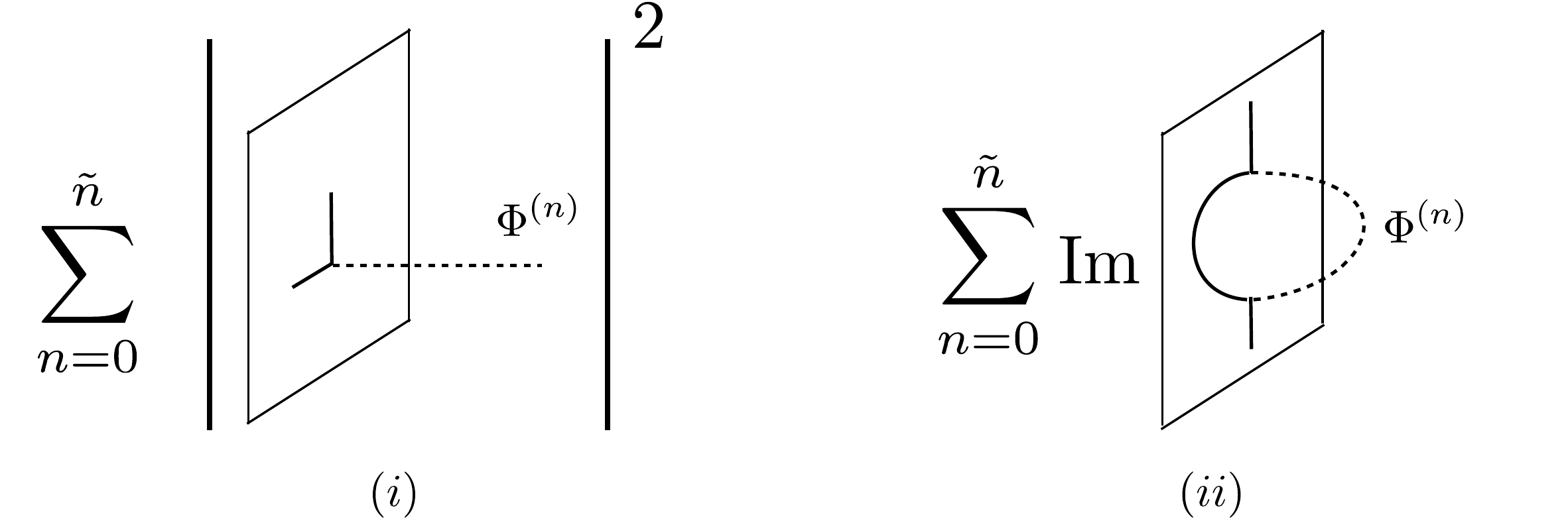} \\
\caption{ 
Two typical processes involving KK sums. \textit{(i)} Emission of KK modes.
\textit{(ii)} Virtual KK modes exchange. 
\label{fig:UVdiags}
}
\end{figure}

It can be often useful,  at least at the conceptual level, to express the propagator $\Delta_p(z,z')$ in a spectral representation, writing it  as the infinite sum of KK modes
\be
\Delta_p(z,z')= i \sum_{n=0}^{\infty} \frac{f_n(z)f_n(z')}{p^2-m_n^2}\, \label{se:KK_dec}
\ee
where  the $f_n(z)$ are the orthonormal KK wavefunctions.
The KK decomposition provides a convenient 4d viewpoint on the 5d physics.~\footnote{
An important conceptual pitfall is that  the EFT for KK modes breaks down in the IR region, 
see Sec.~\ref{se:EFT}.   } However, in the amplitudes, unless one is interested in one or a few KK modes, one ends up with sums involving the KK masses $m_n$ as well as the profiles $f_n(z)$, whose evaluation can be quite challenging.  Here we develop a systematic trick to evaluate the KK sums.   

Sums over KK modes appear either in  amplitudes with virtual KK modes or in square amplitudes with KK modes in the external legs.  Our trick applies to both cases. The general idea is the following. Let us consider a generic quantity
\be
X= \int dz \int dz' \sum_{n=0}^{\tilde n} f_n(z)f_n(z')  A(m^2_n;z,z') \,, \label{eq:X1}
\ee
which describes either an amplitude with KK  modes in an internal  line or a square amplitudes with KK modes in an external line. We can rigorously re-express $X$ using the propagator and a contour integral enclosing the poles up to $ \tilde n$, 
\be
X = -\frac{1}{2\pi } \int dz \int dz'  \oint_{{\cal C}[\tilde n]} d\rho \Delta_{\sqrt{\rho}}(z,z') A(\rho; z,z') \,.
\label{eq:X2}
\ee
The contour ${\cal C}[\tilde n]$ is taken counterclockwise. Clearly, putting Eq.~\eqref{se:KK_dec} in Eq.~\eqref{eq:X2} and using the residue theorem   gives back Eq.~\eqref{eq:X1}. 
The interest of Eq.~\eqref{eq:X2} is that we can now    use the closed-form expression of the propagator Eq.~\eqref{eq:propgen}, or its various limits described in Sec.~\ref{se:lims}.

As a simple example, one can consider the creation of real KK modes  from an interaction localized on the  UV brane. The coupling of a given KK mode $n$ to the UV brane is proportional to $f_n(z_0)$, and the  emission rate for the individual mode takes the form
\be
\Gamma_n=A(m_n)\left(f_n(z_0)\right)^2 
\ee
where $A(m_n)$ depends on $m_n$ at least via a phase space factor. The emission  of \textit{any} KK mode is given by 
\be
\Gamma=\sum_{n=0}^{\tilde{n}} \Gamma_n= \sum_{n=0}^{\tilde{n}}A(m_n)\left(f_n(z_0)\right)^2 \,,
\ee
where the heavier mode $\tilde{n}$ is set by the kinematic threshold encoded in $A(m_n)$. The contour integral trick then gives 
\be
\Gamma=-\frac{1}{2\pi} \oint_{{\cal C}[\tilde n]} d\rho \Delta_{\sqrt{\rho}}(z_0,z_0) A(\sqrt{\rho}) \,.
\ee
By the optical theorem, this quantity gives also the imaginary part of the diagram with an internal $\Phi$ line starting and ending on the UV brane.   
Both quantities are represented in Fig.~\ref{fig:UVdiags}.

\begin{figure}
\center
\includegraphics[width=13 cm,trim={3cm 6cm 3cm 6.5cm},clip]{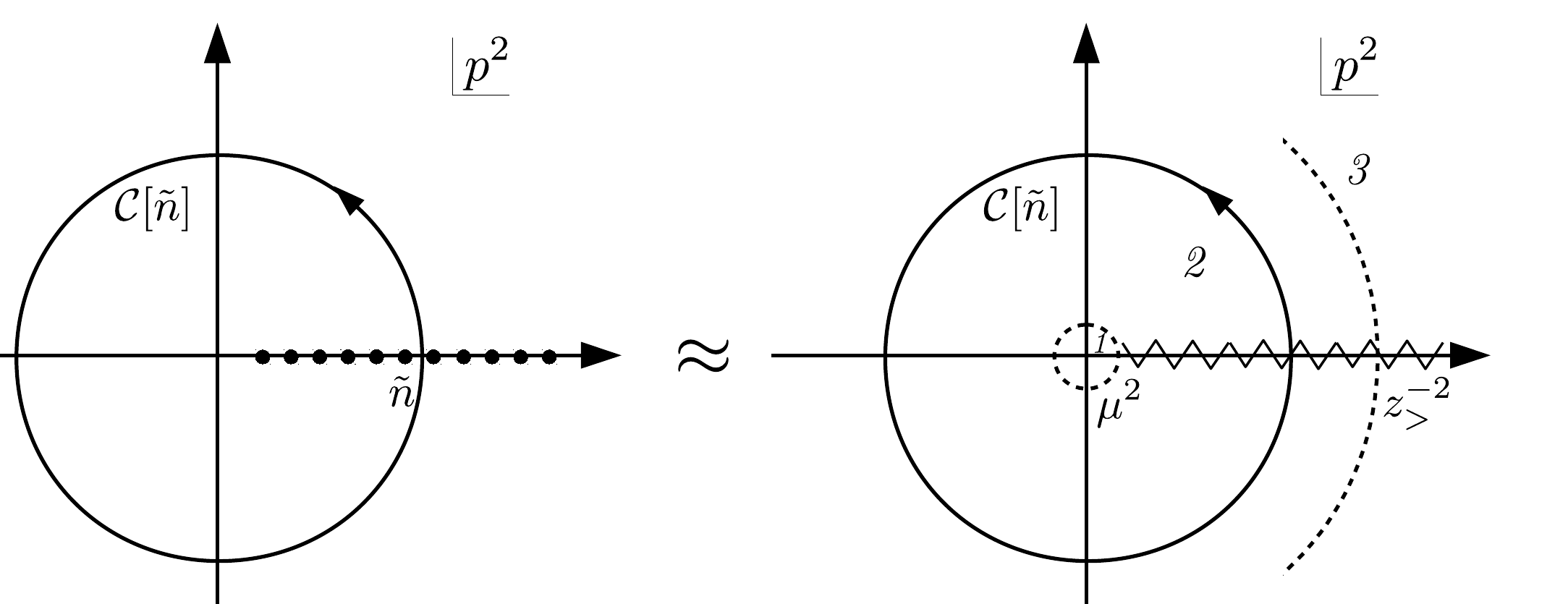} \\
\caption{
Contour integration in the $p^2$ complex plane. The infinite series of poles from the KK representation of the  propagator is shown on the left-hand side and  the branch cut from its closed form is shown on the right-hand side.  The approximation  regions are also pictured. 
\label{fig:Contours}
}
\end{figure}

Let us briefly turn to a calculation in AdS. 
A sum which we will be of use in many of the subsequent calculations is
\be
\sum_{n=0}^{\tilde n}f_n(z) f_n(z')=- 
 \frac{1}{2 \pi }
 \oint_{{\cal C}[\tilde n]} d\rho \Delta_{\sqrt{\rho}}(z,z')\,, \label{eq:ff_int}
\ee
with the $\tilde n$ pole lying in region \textit{2} of position-momentum space. In the corresponding limit Eq.~\eqref{eq:propaII+}, the first term is analytic and does not contribute to the integral. The non-analytic term in Eq.~\eqref{eq:propaII+} does contribute, because it has a branch cut. In  the $\rho$ variable, the branch cut is along the real axis, and the contour ${\cal C}[\tilde n]$ must cross it since it has to enclose the poles, thereby picking the discontinuity across the real axis. The integration contour and the properties of the propagators are shown in Fig.~\ref{fig:Contours}.

Choosing the ${\cal C}[\tilde n]$ contour to be a circle with radius $A$ enclosing the poles up to $\tilde n$, \textit{i.e.} ${\cal C}[\tilde n]=\left(Ae^{i\theta}|\theta\in]-2\pi, 0]\right) $ with   $m^2_{\tilde n}<A<m^2_{\tilde n +1}$, the integral is found to be\footnote{The choice of domain of $\theta$ has to do with conventions for Bessel functions, but one can simply check that it is this  domain which gives a physical result.  }
\begin{align}
\sum_{n=0}^{\tilde n}f_n(z) f_n(z')\approx  & \nonumber \\
\frac{2k\,(kz)^2(kz')^2}{\Gamma(\alpha+1)\Gamma(\alpha+2)\,b^2_{\rm UV}}
&\left(\frac{b_{\rm UV}+2\alpha }{(z_>k)^\alpha}
-b_{\rm UV} (z_>k)^\alpha
 \right) 
\left(\frac{b_{\rm UV}+2\beta }{(z_<k)^\alpha}
-b_{\rm UV} (z_<k)^\alpha
 \right) 
{\left( \frac{m_{\tilde n}}{2k }\right)^{2\alpha+2}}\,.
\label{eq:int_1}
\end{align}
This expression is approximate since one has used the limit Eq.~\eqref{eq:propaII+} of the propagator. Also, one has approximated the radius to $m_{\tilde n}$ because the spacing between modes is $O(\mu)$ and is negligible at our level of approximation, implying that the KK modes are essentially treated as a continuum.   The result of Eq.~\eqref{eq:int_1} can be numerically  compared to the exact expression,  and one finds that the agreement remains of $O(1)$ up to $m_{\tilde n}$ of order of a few $1/z_>$, beyond which one enters too much in region \textit{3} where the approximation breaks down. 

This integral is very useful because it often appears when phase space factors are neglected.
 It can for instance be applied to the case of KK mode emission discussed above, when the KK modes are away from the kinematic threshold. Conversely these examples are another way to see that the non-analytic term of Eq.~\eqref{eq:propaII+} is the one encoding the light component of the KK continuum. 
 Indeed,  it is clear that only light modes can appear in the processes of Fig.~\ref{fig:UVdiags} because of the kinematic threshold. It turns out that it is the non-analytic term which contributes to these processes, while the contribution from the analytic term vanishes.

\section{Dressed propagator  \label{se:scalar}}

In this section and the next one we investigate the scalar propagator dressed by one-loop 1PI insertions.\,\footnote{Exact loop calculations in AdS are typically difficult, see \textit{e.g.} \cite{Aharony:2016dwx,Giombi:2017hpr}. Our approach  relies instead on approximations, including the limits of Sec.~\ref{se:lims}. }
In full generality, given a 1PI diagram in position-momentum space $i\Pi_p(z,z')$, the dressed propagator satisfies
\be
{\cal D} \Delta_p(z,z') - \frac{1}{\sqrt{\gamma}} \int du\, \Pi_p(z,u) \Delta_p(u,z') = - i \frac{\delta(z-z')}{\sqrt{\gamma}}\,.
\label{eq:EOM_dressed}
\ee
One can for instance verify that this is equivalent to the geometric series representation by treating the insertion perturbatively. Writing $\Delta_p(z,z')=\Delta^{(0)}_p(z,z')+\Delta^{(1)}_p(z,z')+\ldots$, we can note that the equation of motion  $\Delta^{(1)}_p(z,z')$ is sourced by $\int du\, \Pi(z,u) \Delta^{(0)}_p(u,z')$, thus implying \be\Delta^{(1)}_p(z,z') =  i \int du \int dv\, \Delta^{(0)}_p(z,u) \, \Pi_p(u,v)\, \Delta^{(0)}_p(v,z') \,,\ee
which is precisely the term with one $i\Pi$ insertion in the geometric series.~\footnote{For the last step recall that ${\cal D}\Phi=J$ is solved by $\Phi(z)=i \int du \Delta(z,u) J(u)$.} 
 Higher terms of the series are obtained recursively.

 As a warm-up in this section we consider  the dressing of $\Phi$ by a loop of another scalar $\varphi$ with cubic bulk interaction
 \be
{\cal L}_{\rm int} \supset \lambda \Phi\varphi^2 \,,
\ee
where $\lambda=O(\sqrt{k})$. 
The bulk mass parameter for $\varphi$ is denoted $\beta$, with $\beta^2 = \frac{m_\varphi^2}{k^2}+4 $.   
This self-energy diagram is shown in Eq.~\eqref{eq:1PI_scalar}. In terms of Kaluza Klein modes, the diagram involves two sums since there are two internal lines of $\varphi$. For our purposes, we are only interested in the imaginary part of $\Pi$.  This means that the KK sums   end at  the kinematic threshold for production of real KK particles, $p=m_n+m_{n'}$. 
Applying the contour integral trick shown in Sec.~\ref{se:sums}, the imaginary part of the diagram is found to be \begin{align}
& {\rm Im} \bigg[ \frac{1}{i}\adjustbox{raise =-1cm}{ \includegraphics[width=2. cm,trim={0.9cm 0.4cm 0.8cm 0cm},clip]{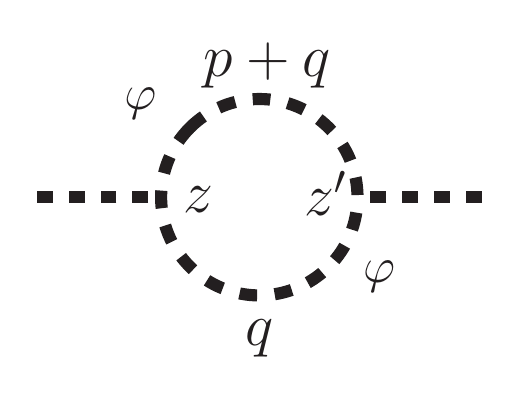}
} \bigg] = {\rm Im} \,\Pi_p(z,z')=
 \nonumber
\\
&\frac{1}{8 \pi^2} \frac{\lambda^2}{(kz)^5(kz')^5} \oint_{{\cal C}[p]} d\rho \oint_{{\cal C}[p-\sqrt{\rho}]}d \rho' \Delta_{\sqrt{\rho}}(z,z')\Delta_{\sqrt{\rho'}}(z,z')
{\rm Im}\left[\frac{1}{i} \int\frac{dq^4}{(2\pi)^4} \frac{1}{(p^2-\rho)((p+q)^2-\rho')}
 \right]\,. \label{eq:1PI_scalar}
\end{align}

Our aim is to study  the effect of this insertion on the behaviour of the propagator. More precisely our focus 
is on region \textit{3} (\textit{i.e.} $1/z_> \lesssim |p| \lesssim 1/z_<$).
Since in this regime  we have $p \gtrsim \frac{1}{z_>}$,  the values of $\sqrt{\rho}, \sqrt{\rho}'$ can lie anywhere with respect to the inverse coordinates of the self-energy,  \textit{i.e.}  all the regions of position-momentum space from \textit{1} to \textit{4} are in principle accessible to the propagators inside the loop. However there are no contributions from propagators lying in region \textit{1} since they are analytic in this region. Contributions from regions \textit{2} to \textit{4} are non-analytic and are thus in principle non-zero.

To proceed in a convenient way, we \textit{hypothesize} that  propagators are actually  suppressed in the region $p \gtrsim 1/z_>$. This hypothesis will be checked at the end of the calculation. 
 When assuming it is true, it implies that the propagators in the loop are suppressed for $\sqrt{\rho},\sqrt{\rho}' \gtrsim 1/z_>$, as a result of their own dressing. This implies that the KK sum is truncated at $\sim 1/z_>$, \textit{i.e.} the contour integral is over a circle of radius $\sim 1/z_>$.   
 It  follows that the contributions to the self-energy  come  from propagators lying in regions  \textit{2}, while 
  contributions from  regions \textit{3}, \textit{4} are exponentially suppressed. Following this approach, if the hypothesis of suppression for $p \gtrsim 1/z_>$ is verified, 
   we will only need to  calculate the contribution from the non-analytic part of the propagator in region \textit{2}. For concreteness we take the integration contour with a radius $|\rho|=1/z^2_>$,  this contour is noted  ${\cal C}[z_>^{-1}]$.

The imaginary part of the bracket in Eq.~\eqref{eq:1PI_scalar} is a standard 4d result. The imaginary part of the self-energy is found to be
\be
{\rm Im}\, \Pi_p(z,z') \approx \frac{1}{128 \pi^3} \frac{\lambda^2}{(kz)^5(kz')^5}  \oint_{{\cal C}[z_>^{-1}]} d\rho \oint_{{\cal C}[z_>^{-1}]}d \rho' \Delta_{\sqrt{\rho}}(z,z')\Delta_{\sqrt{\rho'}}(z,z')
\frac{K(p^2;\rho,\rho' )}{p^2}
\, 
\ee 
where 
\begin{align}
K(p^2;\rho, \rho' ) =
\frac{\sqrt{(p^2-(\sqrt{\rho}+\sqrt{\rho}')^2)(p^2-(\sqrt{\rho}-\sqrt{ \rho}')^2)}}{p^2}
\end{align}
is the familiar 2-body kinematic factor.
 Since we are assuming we are in the $p>1/z_>$ region, the values of $\sqrt{\rho}$, $\sqrt{\rho}'$ are small with respect to $|p|$ and we can safely  take $K\approx K(p^2;0,0)=1$.  
The contour integrals now reduce to the one already  calculated in Sec.~\ref{se:sums}, see Eqs.~\eqref{eq:ff_int},~\eqref{eq:int_1}). For endpoints away from the UV brane the dependence on $b_{\rm UV}$ disappears (this can be seen in Eq.~\eqref{eq:propaII+}). We show only this case for simplicity, the general case  works similarly.   The final result is found to be
\be
{\rm Im}\Pi_p(z,z')\approx
 \frac{1}{8\pi}
\frac{\lambda^2\,k^2}{\Gamma^2(\beta+1)\Gamma^2(\beta+2)}
 \frac{1}{ (kz)^3 (kz')^3}  
\left( \frac{z_<}{4 z_>}\right)^{2\beta+2}
\,. \label{eq:ImPi_scal}
\ee

At that point we have a simple expression for the 1PI insertion. However it is still non local and it is thus difficult to solve the dressed equation of motion. To go further we shall use a position space version of the narrow width approximation (NWA).  The position space NWA amounts to a $\partial_5$ expansion of $\Pi$ where the $\partial_5$ derivatives act on the propagator.~\footnote{This 5d NWA should not be confused with a 4d NWA for individual KK modes, which often does not hold for heavy KK modes.} 
 It can be equivalently seen as an expansion over the basis of the Dirac delta's derivatives, 
 \be
 \Pi_p(z,z')=F_0(z) \delta(z-z')- F_1(z)\delta'(z-z')+\frac{F_2(z)}{2}\delta''(z-z')+\ldots\,. \label{eq:Pi_exp}
 \ee
This can be directly obtained from the dressed equation of motion Eq.~\eqref{eq:EOM_dressed}, where $\Pi_p$ is convoluted with $\Delta_p$. The coefficients of the expansion of the $\Pi_p$ distribution in Dirac Delta's derivative are found to be given by the moments of the distribution, \footnote{
This can be guessed by direct computation, or can be found by thinking in terms of the characteristic function of  $\Pi_p(z,z')$ in Fourier space. }
\be 
F_n(z)=\int dz' (z-z')^n  \Pi_p(z,z') \,.
\ee 
With the NWA expansion Eq.~\eqref{eq:Pi_exp}, $\Pi_p(z,z')$ is decomposed as a sum of local operators and the equation of motion simplifies to an ordinary differential equation. This is a simplification in itself, however this trick is truely appealing because of the property of dressing:
 the local terms can be consistently included step by step in the equation of motion, provided one determines the dressed propagator at every step. 
 For instance instead of solving in the presence of the two first terms, one can solve for the first term, then dress the obtained solution with the second term. For our purposes we will not need to go further than the quadratic order.

The three first coefficients $F_n(z)$ are given by 
\be
F_0(z)=\lambda^2kC_0  \,
 \frac{1}{(k z)^5} \,, \quad 
 F_1(z)= \lambda^2 C_1\,
 \frac{1}{(k z)^4}  \,, \quad
 F_2(z)=\frac{\lambda^2}{k} C_2\,
 \frac{1}{  (k z)^3}  \,,
\ee
with
\be
\begin{pmatrix}
C_0 \\ C_1 \\ C_2
\end{pmatrix}
=
\frac{1}{128\pi}
\frac{1}{16^{\beta}\Gamma^2(\beta+1)\Gamma^2(\beta+2)}
\begin{pmatrix}
\frac{1}{2\beta}+\frac{1}{2\beta+4} \\
\frac{1}{2\beta+1}+\frac{1}{2\beta+3} -\frac{1}{2\beta}-\frac{1}{2\beta+4} \\
\frac{2}{2\beta+2}+\frac{1}{2\beta}+\frac{1}{2\beta+4}-\frac{2}{2\beta+1}
-\frac{2}{2\beta+3}
\end{pmatrix}
\ee

Let us investigate how  these local operators deform the free solutions (given by Eq.~\eqref{eq:solsAdS}). For our purposes it is enough to consider the effect of each of them separately. 
The first term of the expansion  $\Pi(z,z') \supset F_0(z) \delta(z-z') $, when included in the dressed equation of motion Eq.~\eqref{eq:EOM_dressed}, gives a contribution whose $z$-dependence is the same as the bulk mass term. Therefore, at leading order int the NWA, the bulk mass develops a (negative) imaginary part. This is similar to what happens to a propagator in flat space. 
Defining
\be \tilde \alpha= \sqrt{\alpha^2-i C_0 \lambda^2/k} \,,
\ee
with $\alpha$ real, the solutions to the homogeneous equation of motion become
\be
z^2 J_{\tilde\alpha}(pz)\,,\,\,z^2 Y_{\tilde\alpha}(pz)\,
\label{eq:solsAdSC0}
\ee
instead of the free solution Eq.~\eqref{eq:solsAdS}. 
Having a slightly complex order in the Bessel functions implies that the poles are slightly shifted away from the real axis by a constant amount.  
That the bulk mass develops an imaginary part because of dressing is certainly a noteworthy feature, and is after all what we expect from a 5d theory. However this feature  is not very relevant for our purpose, because the behaviour of the propagator is not critically modified and in particular no suppression happens when  the propagator lies in region \textit{3} or \textit{4}.

The second term of the expansion $\Pi(z,z') \supset F_1(z) \delta'(z-z') $ induces a $\partial_5$ derivative and has the right $z$ factor to change the coefficient of the $\partial_5$ term in the ${\cal D}$ operator. In the presence of this term the solutions are deformed to 
\be
z^{\gamma} J_{\scriptscriptstyle \sqrt{m_\Phi^2+\gamma^2}}(p z)\,,\,\,z^{\gamma} Y_{\scriptscriptstyle \sqrt{m_\Phi^2+\gamma^2}}(pz)\,
\label{eq:solsAdSC1}
\ee
where $\gamma=2-i C_1\lambda^2/2k$.
We can see that this deformation gives again a complex contribution to the order of the Bessel function. It also induces a $z$-dependent phase. But again, these deformations to the free solution are small effects which are not too relevant for our study.

The third insertion $\Pi(z,z') \supset F_2(z) \delta''(z-z') $ modifies the $\partial_5^2$ term of the $\cal D$ operator and gives solutions where, in top of a complex Bessel function order and $z$-dependent phase, the argument of the Bessel function is also changed. Showing only this last effect for simplicity, the solutions take the form
\be
z^2J_{\alpha}\left(\frac{pz}{\sqrt{1+i C_2 \lambda^2/2k}} \right)\,,\,\, z^2Y_{\alpha}\left(\frac{pz}{\sqrt{1+i C_2 \lambda^2/2k}} \right) \,. \label{eq:solC2}
\ee
This deformation is the  important one, because it changes the phase of the Bessel's function argument. As a consequence, there are no poles along the real axis, and the Bessel functions rather  have an exponential behaviour controlled by the imaginary part of the argument. 
The full propagator in the presence of this deformation is given by the free propagator Eq.~\eqref{eq:propa_gen} where $p$ is replaced by $p/ \sqrt{1+i C_2 \lambda^2/2k}\approx p (1-i C_2 \lambda^2/4k)$. In the $p\gtrsim 1/z_>$ region, for timelike momentum the propagator behaves therefore as
\be
\Delta_p\left(z,z'\right) \propto e^{- C_2 \lambda^2/4k \,p  z_>}\,. \label{eq:Delta_exp}
\ee 
From this result we conclude that bulk interactions induce an exponential suppression of the propagator: the IR region of the Lorentzian AdS background is opaque. 
This is a quantum effect, unlike the case of spacelike momentum where suppression occurs in the free propagator, the suppression is here controlled by the interaction-dependent,  loop-induced parameter $C_2 \lambda^2/4k$.

  A number of points require discussion to ensure that our result is solid.

First, regarding the NWA expansion, the higher order terms do not have a qualitative impact on the exponential behaviour found in Eq.~\eqref{eq:Delta_exp}, precisely  because the propagator is already exponentially suppressed in the region of interest. Concretely, insertions with higher powers of $\partial_5$ acting on Eq.~\eqref{eq:Delta_exp} bring powers of $(C_2 \lambda^2/4k) pz $ but do not change the exponential suppression, which is thus expected to be a robust feature.~\footnote{ We may also note that the coefficients  of the expansion have a mild hierarchy, with for instance  $C_2/2C_0\sim 1/13 $ for $\beta=1$, $C_2/2C_0\sim 1/60 $ for $\beta=3$, which somewhat reduces the higher terms. }

Second, let us come back to the hypothesis
 that the  propagators in the loop are suppressed outside of region \textit{2},   such that only region \textit{2} contributes to the loop integral.   Since we  indeed find exponential suppression, the  hypothesis is validated.    This basic calculation is sufficient for conceptual discussion, but  if one wishes to be more precise, 
 one should take into account that we have truncated the mode sums at  $p\approx 1/z_>$ (see Eq.~\eqref{eq:1PI_scalar}),  while the suppression found in Eq.~\eqref{eq:Delta_exp} occurs more precisely at  $p\sim ( 4k/  C_2 \lambda^2) /z_>  $.  The truncation of the loop should match this value for the calculation to be more accurate.
  This is a self-consistent problem which can be solved by  iterating the calculation until reaching a fixed point.  For instance a self-consistent result assuming 5d strong coupling \cite{Chacko:1999hg, Agashe:2007zd} is 
 \be
 \lambda= 24^{1/3} \pi \sqrt{k}\,,\quad \beta=1\,,\quad z_>\lesssim \frac{2.8} {p}\, \label{eq:pointscal}
 \ee
where the last inequality describes the truncation of the loop integral.  

Notice that this is at the limit of validity of region \textit{2}. 
 For smaller $\lambda$ or larger $\beta$, $C_2$ gets smaller and there should be KK modes contributing beyond region \textit{2}. The contributions from region \textit{3},  \textit{4} are however more difficult to evaluate analytically.    Going in that direction, one may as well compute the dressed propagators numerically for  given $p$, $z$, $z'$.  This is a heavy task which is beyond the scope of this work; a mere estimate is enough for the present study. Our estimate uses only region \textit{2} and is thus conservative since  it selects at worse a subset of the KK modes running in the loop.

Third, one may wonder about the magnitude of higher-order loops contributions. The relative magnitude of one-loop and higher-order loop diagrams depends on $p$ and $z$. Namely, the higher-order contributions  tend to become non-negligible in the IR, at some point inside the  $pz_> >1$ region.  
This is expected, since we are in an effective theory with non-renormalizable interactions. As a matter of fact, one approach to determine the validity of an EFT is precisely to compare loops of different order---the EFT cutoff being then the region of position-momentum space at which loops of different order become of same  magnitude.  Generally, because of loop suppression, one expects the higher-order loops to become relevant when the exponential suppression from one-loop is already established, such that no crucial qualitative change is expected from higher-loop diagrams.

Let us for instance focus on a two-loop self-energy diagram, proportional to $\lambda^4$. 
The imaginary part of the two-loop diagram is tied to the 3-body decay calculated in Sec.~\ref{se:decay} by the optical theorem. An evaluation in the $p<1/z_>$ regime is given in Eq.~\eqref{eq:recursion}. It turns out that even for the point Eq.~\eqref{eq:pointscal} the two-loop contribution is suppressed by two orders of magnitude with respect to the one-loop contribution. This implies that $p$ has to be sensibly larger than $1/z_>$ for  the 2-loop contribution to become sizable, and  the exponential suppression from one-loop is thus already established when this happens.

This concludes our  calculation of dressing. The emphasis in this section has been on the conceptual steps  and on the approximations taken, considering the dressing  from a simple scalar loop.   In next section we  apply the same approach to the gravity case, the steps followed are essentially the same but the  calculations are slightly more technical.

\section{Gravitational dressing \label{se:grav}}

Five-dimensional gravity is always present in the theory, hence 5d gravitons induce  a universal contribution to the dressing of matter fields. 
The gravity formalism is somewhat heavier than the scalar case  because each KK graviton has five degrees of freedom, two of helicity-two, two of helicity-one, one of helicity-zero, each with their own coupling to the stress-energy tensor. The trace of the metric is a non-physical, ghosty degree of freedom which also contributes to the loop. 
We  closely follow  the formalism of \cite{Dudas:2012mv}. \footnote{To the exception that we use the opposite metric signature.}

The canonically normalized metric fluctuation  around the AdS background is defined by
\be
g_{MN}=\gamma_{MN}+\sqrt{\frac{2}{M^3_*}} h_{MN}\,.
\ee
The expansion of the gravity action Eq.~\eqref{eq:5d_action}  up to quadratic order is well-known (see \textit{e.g.}\cite{Boos:2002hf,Hinterbichler:2011tt}), giving the action for the 5d graviton $h_{MN}$, 
\begin{align}
& S^h=  & \\ \int dX^M & \sqrt{\gamma} \,\bigg( 
\frac{1}{2}\nabla_R h_{MN}\nabla^R h^{MN} - 
\frac{1}{2}\nabla_R h\nabla^R h+
\nabla_M h^{MN}\nabla_N h
-  \nabla_M h^{MN} \nabla^R h_{RN}
\\ & \quad\quad\quad +\frac{1}{2}(h^2_{MN}+h^2) \nonumber 
 + \sqrt{\frac{1}{2 M^3_*}}h^{MN} T_{MN} \bigg) \,
\nonumber 
\end{align}
where
\be
T_{MN} = 2\frac{\delta {\cal L}_{\Phi}}{\delta \gamma_{MN}} - \gamma_{MN}{\cal L}_{\Phi} \,.
\ee

Following \cite{Dudas:2012mv},  all the degrees of freedom can be disentangled using field redefinitions and Faddeev-Popov gauge fixing. 
Defining $\hat h_{MN}=(kz)^2 h_{MN}$  then splitting the graviton components as
\be
 \tilde h_{\mu\nu}=\hat h_{\mu\nu}-\frac{1}{4}\eta_{\mu\nu}h^\rho_\rho \,, \quad 
  B_\mu = \frac{\sqrt{2}}{kz}\hat h_{\mu 5}\,,\quad 
 \chi=\frac{1}{2}\left(\hat h_\mu^\mu-2 \hat h_{55}\right)\,, \quad
  \phi=\frac{\sqrt{3}}{\sqrt{2}(kz)^2}\hat h_{55}\,,
  \ee
and defining the sources 
\be
\tilde T_{\mu\nu}=T_{\mu\nu}-\frac{1}{4}\eta_{\mu\nu}T_\rho^\rho\,,\quad
\tilde T_{55}=T_{55}+\frac{1}{2}T_\rho^\rho\,, 
\ee
the graviton action $S^h=\int dx^M {\cal L}^h$ takes the simple form
\begin{align}
{\cal L}^h= & 
\frac{1}{2}\bigg(
\frac{1}{(kz)^3}(\partial_R \tilde h_{\mu\nu})^2+\frac{1}{kz}(\partial_R  B_{\mu})^2
+ k z (\partial_R \phi)^2 -\frac{1}{(kz)^3} (\partial_R \chi)^2
\bigg)  \label{eq:Lagh}
 \\ &
+\frac{1}{\sqrt{M_*^3}}
\left(
\frac{1}{\sqrt{2}(kz)^3} \tilde h^{\mu\nu}\tilde T_{\mu\nu}
+
\frac{1}{(kz)^2} B^\mu T_{\mu 5}
+\frac{1}{2\sqrt{2}(kz)^3} \chi T_\mu^\mu
+\frac{1}{\sqrt{3}kz} \phi \tilde T_{55}
\right)\,. \nonumber
\end{align}
In Eq.~\eqref{eq:Lagh} all contractions are done with the Minkowski metric. The zero mode of $\tilde h_{\mu\nu}$ matches the 4d graviton in the 4d low-energy EFT. The zero mode of $\phi$ corresponds to the radion field. As expected the $\chi$ field has ``wrong-sign'' kinetic term and is an unphysical degree of freedom.

The Feynman  propagators in position-momentum space for the degrees of freedom $(\tilde h_{\mu\nu},B_\mu, \phi,\chi)$ are expressed in terms of the scalar propagator by
\begin{align}
\left\langle \tilde h_{\mu\nu}(p,z) \tilde h_{\rho\sigma}(-p,z') \right\rangle & = \eta_{\mu\nu}\eta_{\rho\sigma} \Delta_p(z,z') \big|_{\alpha=2,b_i=0}\,, \quad \\
\left\langle  B_{\mu}(p,z)  B_{\nu}(-p,z') \right\rangle & =
\eta_{\mu\nu} \Delta_p(z,z') \big|_{\alpha=1,b_i=\infty}\,, \quad \\
\left\langle  \phi(p,z) \phi(-p,z') \right\rangle & = \Delta_p \big|_{\alpha=0,b_i=0}\,,\quad \\
\left\langle  \chi(p,z) \chi(-p,z') \right\rangle& =  -\Delta_p \big|_{\alpha=2,b_i=0}\,.
 \end{align}
 The value $b_i=\infty$ means that  boundary conditions are Dirichlet for the vector component.  

The stress-energy tensor in our study is the one of the bulk scalar $\Phi$,
 \be
T_{MN}= \partial_M \Phi \partial_N \Phi - \frac{1}{2}\gamma_{MN} \left(\partial_M \Phi \partial^M \Phi  -M^2_\phi \Phi^2 
 \right)\,.
\ee    
The sources for the disentangled graviton components are thus
   \be
   \tilde T_{\mu\nu} = \partial_\mu \Phi\partial_\nu \Phi - \frac{1}{4}\eta_{\mu\nu} (\partial_\mu \Phi)^2 \,,
   \ee
      \be
   T_{\mu}^\mu=- (\partial_\mu \Phi)^2+2(\partial_5 \Phi)^2+\frac{2}{(kz)^2}M^2 \Phi^2 \,,
   \ee
      \be
   \tilde T_{55}=\frac{3}{2}(\partial_5 \Phi)^2+\frac{1}{2\,(kz)^2}M^2 \Phi^2 \,,
   \ee
      \be
   T_{\mu 5}=\partial_\mu \Phi \partial_5 \Phi \,.
   \ee
Having all the Feynman rules needed, we can now calculate the imaginary part of the self-energy
\be
{\rm Im} \,\Pi_p(z,z')=
{\rm Im} \,\Pi^{h}_p(z,z')+
{\rm Im} \,\Pi^{B}_p(z,z')+
{\rm Im} \,\Pi^{ \phi}_p(z,z')+
{\rm Im} \,\Pi^{\chi}_p(z,z') \,,
\ee
\be
{\rm Im} \,\Pi^{h,B,\phi,\chi}_p(z,z')={\rm Im} \bigg[ \frac{1}{i}\adjustbox{raise =-0.3cm}{ \includegraphics[width=2.6 cm,trim={0.cm 0.8cm 0.cm 0cm},clip]{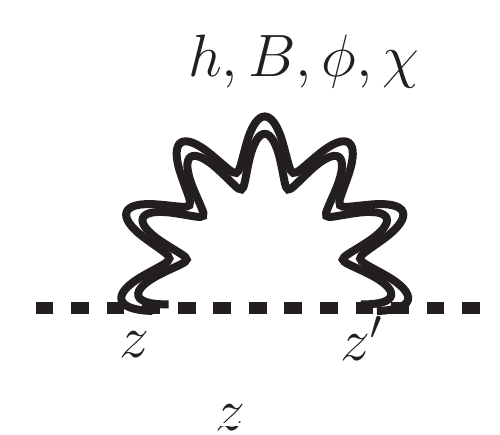}
} \bigg]  \,. \label{eq:ImPigrav}
\ee

Our calculation being in position-momentum space, the vertices from the stress-energy tensor contain  $\partial_5$ derivatives acting on the external $\Phi$ propagators. For our purposes it is convenient to  integrate by part in $z$ for each insertion of $\Pi_p(z,z')$ in the geometric series,  such that these $\partial_5$ derivatives act instead on all propagators and metric factors inside the loop. 
As in previous section we focus for simplicity on endpoints lying away for the UV brane. Hence the UV brane  terms and the  brane contributions generated when integrating by part are irrelevant. Elements of the graviton loop calculation are given in App.~\ref{app:grav_loop}. 

Introducing the 5d gravity coupling strength $\kappa \equiv \sqrt{k^3/M_*^3}=k/M_{Pl}$, the final result is found to be
\be
{\rm Im}\, \Pi^h(z,z')+{\rm Im}\, \Pi^\chi(z,z') = \frac{\kappa^2\,k\, p^2}{64\pi}
\frac{2+3\alpha}{\Gamma(1+\alpha)\Gamma(2+\alpha)}
\frac{1}{(k z_>)^8 }
\left(\frac{z_<}{4z_>}\right)^{\alpha-1}\,, \label{eq:ImPih}
\ee
\be
{\rm Im}\, \Pi^B(z,z') = 
\frac{\kappa^2\,k\, p^2}{32\pi}
\frac{(2+\alpha)(3+2\alpha)}{\Gamma(1+\alpha)\Gamma(2+\alpha)}
\frac{1}{(k z_>)^6 }
\left(\frac{z_<}{4z_>}\right)^{\alpha+1}\,, \label{eq:ImPiB}
\ee
\be
{\rm Im}\, \Pi^\phi(z,z') = 
\frac{\kappa^2\,k^3}{192\pi}
\frac{(2+\alpha)(2+6\alpha+\alpha^2)}{\Gamma^2(1+\alpha)}
\frac{1}{(k z_>)^6 }
\left(\frac{z_<}{4z_>}\right)^{\alpha-1}\,. \label{eq:ImPiphi}
\ee
It turns out that the $h$, $\chi$, $B$ contributions are small with respect to the contributions from $\phi$. They are both suppressed by a factor of $p^2/k^2$, and the $h+\chi$ contribution has also an an extra suppression in $1/(kz_>)^2$.  
In contrast, the contribution from the $\phi$ degree of freedom is  similar to the scalar case described in previous section and can be treated the same way. 

Assuming $\alpha>1$, one can use the position-space narrow width approximation
\be
\Pi^\phi(z,z')=\delta(z-z')F^{\phi}_0(z) + F^{\phi}_1(z)\delta'(z-z')+\frac{F^{\phi}_2(z)}{2}\delta''(z-z') +\ldots
\ee
\be
F^{\phi}_0(z)=\kappa^2\,k^2\,C^{\phi}_0\frac{1}{(kz)^5} \,,\quad
F^{\phi}_1(z)=\kappa^2\,k\,C^{\phi}_1\frac{1}{(kz)^4} \,,\quad
F^{\phi}_2(z)=\kappa^2\,C^{\phi}_2\frac{1}{(kz)^3} \,
\ee
with
\be
\begin{pmatrix}
 C^\phi_0 \\ C^\phi_1 \\ C^\phi_2
\end{pmatrix}=
\frac{1}{48\pi}
\frac{(2+\alpha)(2+6\alpha+\alpha^2)}{4^\alpha\Gamma^2(1+\alpha)}
\begin{pmatrix}
\frac{1}{\alpha}+\frac{1}{\alpha+4} \\
\frac{1}{\alpha+1}+\frac{1}{\alpha+3}
-\frac{1}{\alpha}-\frac{1}{\alpha+4} \\
\frac{2}{\alpha+2}+\frac{1}{\alpha}+\frac{1}{\alpha+4}-\frac{2}{\alpha+1}
-\frac{2}{\alpha+3} \\
\end{pmatrix} \,.
\ee
Then, as in previous section, the $C^\phi_2$ term turns out to induce an exponential suppression of the propagator in the IR region
\be
\Delta_p\left(z,z'\right) \propto e^{- \kappa^2 C^\phi_2 /4 \,p  z_>}\,. \label{eq:Delta_exp_grav}
\ee 

Having an estimate at strong coupling is instructive. In the case of AdS gravity, strong coupling is at $\kappa=O(1)$ \cite{Agashe:2007zd}. Using $\kappa=1$, a point with self-consistent truncation of the loop (see previous section) is for instance
\be
\kappa=1\,,\quad \alpha=2\,,\quad  z_> \lesssim \frac{2.6} {p}\,. \label{eq:pointgrav}
\ee
Again, as in the scalar cubic interaction case, the truncation lies at the limit of validity of region \textit{2}. 
Weaker values of  $\kappa$  or higher $\alpha$ would require to know the contributions from region \textit{3} and \textit{4} to get  an accurate result,  however this is beyond the scope of the present calculations.

\section{EFT considerations}

\subsection{Censorship of the IR region  \label{se:EFT}  }

In the two previous sections we have found that, at least at strong coupling and for $O(1)$ values of the bulk mass parameter (see the examples Eqs.~\eqref{eq:pointscal},~\eqref{eq:pointgrav}), the opaque region of position-momentum space starts  around  $pz_>=O(1)$.
  Let us  compare this result to  estimates of the region where the 5d EFT breaks down. Following the approach of \cite{Goldberger:2002cz}, one includes a higher dimension bilinear operator in the effective 5d Lagrangian, 
\be
{\cal L}_5\supset \frac{1}{\Lambda^2}\square^2\Phi^2\,.
\label{eq:HDO}
\ee
This operator is certainly expected to dominate the propagator in some region of position-momentum space, signaling that the 5d EFT breaks down. Dressing the 5d propagator with the insertion from Eq.~\eqref{eq:HDO}, we can readily observe that in the $p>1/z_>$ region (\textit{i.e.} region \textit{3}), the $kz\, \partial_5$ from the extra derivatives contribute to extra $\frac{k}{\Lambda}pz_>$ factors in the geometric series resulting from the dressing. 
From this one concludes that the  5d EFT becomes invalid around
\be
\frac{k}{\Lambda}pz_> \sim 1 \label{eq:EFT_region} \,.
\ee
 A similar result was obtained in \cite{Goldberger:2002cz} by projecting the propagator onto a single KK mode. 

An estimate of $\Lambda$ is given by  dimensional analysis \cite{Chacko:1999hg, Agashe:2007zd}, 
 \be
 \Lambda^3 \sim 24 \pi^3\, M_*^3 = 24 \pi^3\, \frac{k^3}{\kappa^2}  \,.
 \label{eq:Lambda}
 \ee
For strongly coupled gravity $\kappa \sim 1$ the region of EFT breaking starts thus at $p \sim 10/z_>$. This is typically inside the ``opaque'' region where exponential suppression occurs.  Hence there is no need to forbid by hand 
the region of 5d EFT breaking, the theory censors this region by itself.
For weaker couplings our estimates are not precise enough to draw a similar conclusion, because contributions from regions \textit{3} and \textit{4} should also be taken into account in the dressing (see Sec.~\ref{se:scalar}). 

It is worth noting that the $\kappa$-dependence of the EFT breaking region Eq.~\eqref{eq:EFT_region} obtained from a naive dimensional analysis  Eq.~\eqref{eq:Lambda} does not match the $\kappa$-dependence of the opaque region induced by gravitational dressing. The former goes as $pz_> \propto 1/\kappa^{2/3}$ while the latter goes as $pz_> \propto 1/\kappa^2$. 
Interestingly this slightly disturbing feature can be avoided if one takes the hypothesis that the higher dimensional operators in the theory are only generated from gravity loops. Indeed the bilinear higher derivative operators of the kind of Eq.~\eqref{eq:HDO} are then generated by contributions from gravitational dressing, \textit{i.e.} from 
the loop diagram shown in Eq.~\eqref{eq:ImPigrav}. In such case, the boundary of the EFT breaking region scales as $pz_> \propto 1/\kappa^2$, just like the opaque region. In fact, 
 the real part of $\Pi_p$ induces effective operators which are automatically of same order of magnitude as the imaginary part of $\Pi_p$.
As a result the 5d EFT breaking region and the opacity region should roughly match each other for any values of coupling and bulk mass, since the two effects originate from the same loop. Somehow, the dressing by ${\rm Im}\,\Pi_p$ renders automatically opaque the region where the effects from ${\rm Re}\,\Pi_p$ would be out of control. 

Finally let us emphasize again that the IR opacity and censorship  behaviours discussed in this work have no equivalent in flat space.  First, unlike in the curved case,  it is  possible to have exact KK parity  in a flat extradimension. In such case the KK modes are stable, hence  the $ \Pi_p(z,z')$ self-energy 
does not develop an imaginary part and no effect of opacity can ever occur. 
Second, even if KK parity is broken, the KK modes couple to bulk fields  with the strength of 4d gravity, hence very weakly. The dressing of a bulk field by KK modes becomes relevant only near the 5d cutoff, once enough real KK modes run in the loops, and thus does not seem to have  interesting consequences.

\subsection{CFT interpretation}

The  interpretation of the opacity effect in  a dual CFT picture is clear and was already noticed in \cite{Gherghetta:2003he} in the spacelike momentum case. 

First let us remind that the presence of the IR brane is interpreted in the CFT dual as a spontaneous breaking of conformal symmetry in the IR, and any field localized towards the IR brane is interpreted as a bound state arising from confinement of the CFT (see \textit{e.g.} \cite{ArkaniHamed:2000ds}). Concretely, the UV-to-UV correlators, which are those needed for holography, interact and possibly mix with the IR-localized fields. 
 At low 4-momentum $|p|\sim \mu$, most KK modes are integrated out, and only  light poles---including those from the IR-localized fields---are in the spectrum, and are  understood as the resonances arising in the confined phase of the CFT.   
In contrast, at momentum $|p|\gg \mu$, IR opacity implies that the effects from  IR-brane-localized degrees of freedom (\textit{e.g.} poles) vanish exponentially from the   UV-to-UV correlators. In fact, the IR brane vanishes from the correlators, which become effectively the same as in pure AdS. 

On the CFT side, the vanishing of the  IR-related effects in the correlators matches the  intuition about bound states of size $\sim 1/\mu$. At momentum above $\mu$, an external probe sees the CFT constituents and does not know about the bound states. The low-energy EFT describing the bound states breaks at a scale somewhat above $\mu$ (say $4\pi \mu$), however these vanish from the amplitudes   via the effect of form factors---which are  implemented by the opacity property of AdS.    
Such observations were essentially done in \cite{Gherghetta:2003he} for spacelike 4-momentum, and the present work ensures they apply also for timelike 4-momentum.


The above considerations on the low-energy EFT of ``CFT bound states'' in the IR involve the absolute 4-momentum $p$ and how it compares to the ``CFT breaking scale'' $\mu$. The meaning of $p^\mu$ is trivial on both CFT and AdS sides. 
It is worth pointing out that another set of conclusions, with apparently no or little interplay with the 
aspect discussed above,  is reached when considering the 5d short distance limit of small $|\Delta X^M|\sim 1/\Lambda$. In this limit, curvature becomes negligible, and the same cutoff as in 5d flat space is reached. The existence of the 5d cutoff, while conceptually simple on the AdS side, has nontrivial implications for the CFT side. A simple example is as follows.  A 5d  mass cannot be larger than $\Lambda$. On the CFT side, this implies that the  corresponding conformal dimension cannot be arbitrary large. Hence validity of the 5d EFT implies a nontrivial bound on the dimension of the CFT operator. 
Such constraint on the CFT side is, to the best of our understanding,  unrelated to the CFT's IR behaviour discussed above.  Here the parameter involved to delineate the limit of the 5d EFT is the 5d energy, whose CFT interpretation is not obvious and has to be carefully determined. 
An analysis along these lines involves identifying the AdS Hamiltonian with the CFT dilatation operator. These aspects have  been thoroughly studied in \cite{Fitzpatrick:2010zm} from both AdS and CFT sides.

\section{KK continuum cascade decay rate  \label{se:decay}}

In the previous sections we have found that opacity of Lorentzian AdS is a consequence of bulk interactions,  and happens 
essentially because KK modes can decay into lighter KK modes. The KK modes can often have a width larger than $\mu$ in which case they tend to form a continuum. 
 The term ``KK modes'' is general  since it applies to  a set of modes with any width, but in the following we will use specifically ``KK continuum''  for a set of dressed KK  modes  in internal lines. 
The AdS KK continuum can  undergo cascade decays and there are various reasons to think  about them in detail.

First, one may wonder whether the IR brane actually remains  invisible to a bulk field with $p\gg \mu$  in the presence of cascade decays. 
Indeed, consider for instance a process where a KK continuum is created on the UV brane, propagates to the IR brane where it decays into IR localized states. As we have seen in previous sections, for $p>\mu$ this process is exponentially suppressed. However the KK continuum could also decay in the bulk, splitting its initial timelike absolute momentum $p$ between the various daughter continuums  produced. Repeating the decay process enough times, the absolute momentum keeps fragmenting and  the final daughter KK modes  end up with a  4-momentum $p_f \sim O(\mu)$. Naively this seems to imply that the probability of the initial continuum   to reach the IR brane is not exponentially suppressed. This observation seems to somewhat contradict the picture of IR opacity we have previously obtained, and  therefore some  investigation is needed.

Second,  the cascade decay of the KK continuum is certainly interesting in its own right,  and especially in the context of the AdS/CFT correspondence. Cascade decays in the dual theory correspond to the decay of excitations in the CFT, which are expected to produce 
soft, spherically distributed  final states with high multiplicity, sometimes called ``soft bombs'', see \textit{e.g.}~\cite{Strassler:2008bv}.  
While qualitative studies have been done on the CFT side, 
not so many studies have been on the AdS side, to the best of our knowledge. A Monte Carlo study has been done in \cite{Csaki:2008dt}  in the KK picture, considering the decay of single KK modes, with a focus on the details of the final state distribution. 
In our analytical calculation, we will recover qualitatively some of these features, but our focus will rather be on the estimation of the total rate of the cascade decay. It is not clear if a same calculation has been tried on the CFT side.

For simplicity we focus on a scalar field decaying via a cubic self-interaction, which is  enough to display the key features of the calculation.  The interaction Lagrangian is  ${\cal L}\supset \lambda \Phi^3$. Other interactions like gravity are assumed to be present such that the exponential suppression  starts around $p\sim 1/z_>$, which will be used  to truncate the space integrals. 

Our main goal is to  estimate the rate of a cascade decay. Importantly, since  the KK modes tend to be  broad, a 4d narrow-width approximation  cannot be applied. This means that the rate  cannot be factored out as production rate times many branching ratios, as would be the case for the cascade decay of  narrow particles. Instead the whole squared matrix element has to be calculated without relying on a narrow width approximation.

For our purposes it is enough to study the rate of a 3-body cascade decay, made of two successive decays induced by the cubic vertex. In terms of KK modes the integrated square matrix element is given by
\be \sum_{m,n,p}\int d\Phi_3  |{\cal M}_{mnp}|^2\,, \label{eq:3bodygen}
\ee
with
\be
{\cal M}_{m,n,p}=\adjustbox{raise =-1cm}{\includegraphics[width=0.2\textwidth,trim={6cm 7.5cm 12cm 6cm},clip]{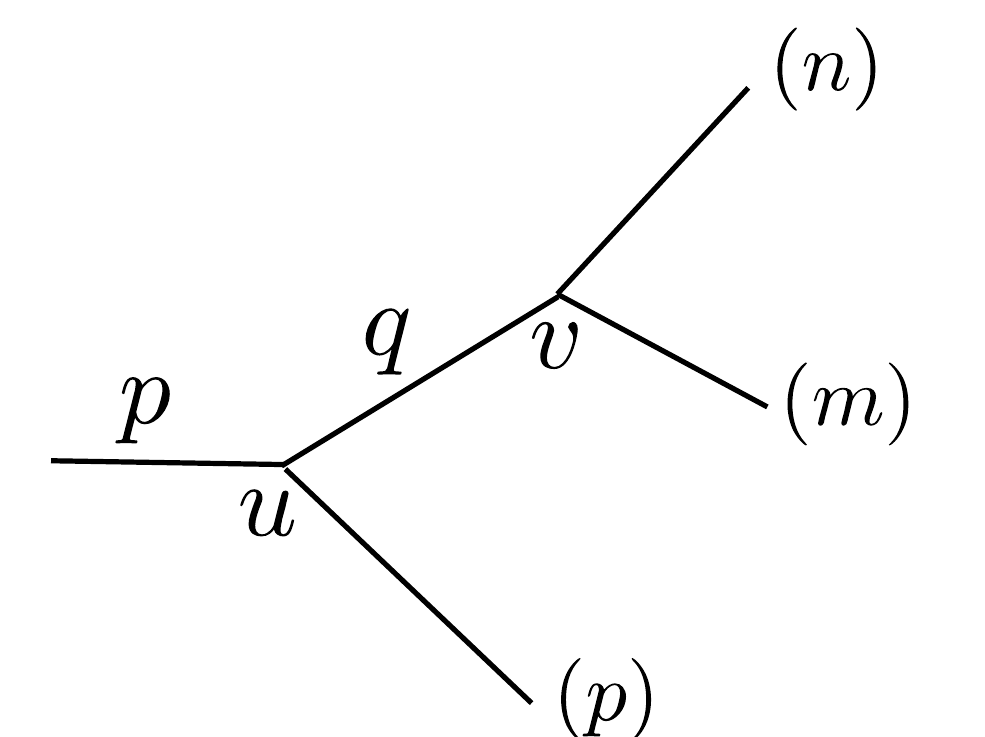}} =\int^{\frac{1}{\mu}}_{\frac{1}{k}} du\int^{\frac{1}{\mu}}_{\frac{1}{k}} dv \frac{\lambda^2}{(ku)^5(kv)^5} \Delta_p(z_0,u)f_{p}(u)
\Delta_q(u,v) f_m(v)f_n(v)\,. \label{eq:3bodygenamp}
\ee
The  exponential falloff of the propagators resulting from dressing---in particular gravitational dressing---is assumed to occur around $p\sim 1/z_>$. The suppression of the propagator amounts to cut the $u$, $v$ position integrals.
Notice  that the EFT of the external KK modes  breaks down in the IR region, for instance for $v\gg1/q$ in Eq.~\eqref{eq:3bodygenamp}, but as discussed in Sec.~\ref{se:EFT} this region is censored by the exponential falloff of the propagators to which the external KK modes are connected.

Let us write the 3-body phase space integral as the sequence of two 2-body decays
\be
\int d\Phi_3(p;p_1,p_2,p_3)= \int \frac{dq^2}{2\pi} \int d\Phi_2(p;q,p_3) \int d\Phi_2(q;p_1,p_2)
\label{eq:Lips3}
\ee
and focus on the secondary decay. In Eq.~\eqref{eq:3bodygen} the expression describing this subdecay  is
\be
\sum_{m,n}\int^{\frac{1}{\mu}}_{\frac{1}{k}} dv \int^{\frac{1}{\mu}}_{\frac{1}{k}} dv' \int d\Phi_2(q;p_1,p_2) \frac{\lambda^2}{(kv)^5(kv')^5} \Delta_q(u,v) \Delta^*_q(u',v') f_m(v)f_m(v')f_n(v)f_n(v')\,.
\ee
Let us first use the contour integral trick on the KK sums and evaluate the phase space integral.  The expression becomes 
\be
\frac{1}{64 \pi^3}\int^{\frac{1}{\mu}}_{\frac{1}{k}} dv \int^{\frac{1}{\mu}}_{\frac{1}{k}} dv' \frac{\lambda^2}{(kv)^5(kv')^5}\Delta_q(u,v) \Delta^*_q(u',v') \oint_{{\cal C}[\sqrt{\rho}]} 
d\rho \oint_{{\cal C}[q-\sqrt{\rho}]}d \tilde \rho
 K(q;\sqrt{\rho},\sqrt{\tilde\rho})   \Delta_{\sqrt{\rho}}(v,v')\Delta_{\sqrt{\tilde\rho}}(v,v')\,.
\ee

Up to this point these are exact expressions, now let us  make approximations. Since the internal propagators are exponentially suppressed because of their dressing, the integration in position-momentum space are restricted to  $u,u', v, v'\lesssim 1/q $, \textit{i.e.} the internal propagators are taken to be non-vanishing only in region \textit{2}. It follows that, because of the kinematic threshold, the $\Delta_{\sqrt{ \rho}}$'s appearing in the final state contour integration are also forced to be in region \textit{2} since $\sqrt{\rho} \leq q$. 
Apart from the restriction of the integration region, the other important approximation we take is to replace the kinematic factor as $K(q;\sqrt{\rho},\sqrt{\tilde\rho})\approx\Theta\left(q-\sqrt{\rho}/2\right)\Theta\left(q-\sqrt{\tilde\rho}/2\right)$.  This is a convenient but 
 rough approximation implying  extra $O(1)$ uncertainties in the subsequent results. Such level of precision is  sufficient for our study.

Given the approximations above, the expression becomes
\be
\frac{1}{64 \pi^3}\int^{\frac{1}{q}}_{\frac{1}{k}} dv \int^{\frac{1}{q}}_{\frac{1}{k}} dv' \frac{\lambda^2}{(kv)^5(kv')^5}\Delta_q(u,v) \Delta^*_q(u',v') \oint_{{\cal C}[q/2]} 
d\rho \oint_{{\cal C}[q/2]}d \tilde \rho \,
    X_{\sqrt{\rho}}(v,v')X_{\sqrt{\tilde\rho}}(v,v') \,
\ee
where we have introduced $X_q(z,z')$,   the non-analytic part of $\Delta_q(z,z')$ in region \textit{2}.
We then integrate over the positions $v$, $v'$. The integration is straightforward but slightly tedious because, as discussed in Sec.~\ref{se:lims}, $\Delta_q$ in region \textit{2} contains several terms. It turns out that the non-analytic term $X_q$  dominates because it has the largest positive power of $v$'s. 
The result reads
\be
\frac{\Gamma^2(-\alpha)}{4096\pi\,4^{6\alpha}(3\alpha+2)^2\Gamma^4(\alpha+1)\Gamma^2(\alpha+2)\,}
u u' (q^2 u u')^\alpha\, \label{eq:int_subdecay}
\ee
while the neglected terms are smaller by at least a factor  $(|q|/k)$.  Contributions from UV brane terms are also negligible. We then observe that Eq.~\eqref{eq:int_subdecay} can be re-expressed as a function of $X_q(u,u')$, 
\be
i\frac{1}{k}\frac{(-1)^{-\alpha}\,\Gamma(-\alpha)}{2048\pi\,4^{5\alpha}(3\alpha+2)^2\Gamma^3(\alpha+1)\Gamma^2(\alpha+2)\,} X_q(u,u')
\,. \label{eq:int_subdecay2}
\ee

Having reduced the expression of the secondary decay, let us return to the full amplitude Eq.~\eqref{eq:3bodygen}. Recall we are  using the  phase space representation of Eq.~\eqref{eq:Lips3}. We can readily use the 2-body phase space approximation on the primary decay, just as done on the secondary decay. Then let us  consider the remaining integral on the $q^2$ variable from Eq.~\eqref{eq:Lips3}. The integrand depends on $q^2$ via Eq.~\eqref{eq:int_subdecay2}.  The integral is along the real axis but  is related to the one over a circle by
$\int_0^{R} dq^2 X_q(u,u')=
\int_{{\cal C}[R]} dq^2 X_q(u,u') \frac{(-1)^{\alpha}}{2i \sin(\pi \alpha)}\,,
$
where $R$ is taken at $p/2$ since one uses the same approximations as in the secondary decay.  As defined in Sec.~\ref{se:sums}, the contour is ${\cal C}[R]=\left(Re^{i\theta}|\theta\in]-2\pi,0]\right)$. 
With these rewritings the $q^2$ integral of $X_q$ has the same form as the contour integral representing a sum of KK modes in final state. This means that we have reduced the integrated 3-body square amplitude into an integrated 2-body square amplitude. Putting all the pieces together, the relation found is
\be 
\includegraphics[width=0.8\textwidth,trim={1cm 7.5cm 3cm 8.9cm},clip]{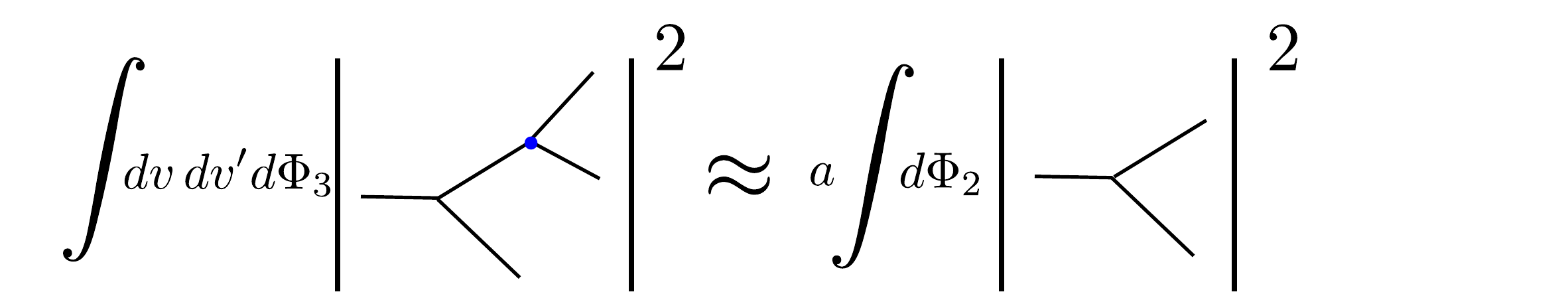} \,
\label{eq:recursion}
\ee
where 
\be
a=\frac{\lambda^2}{k}\frac{1}{4096\pi^2\,4^{5\alpha}(3\alpha+2)^2}\frac{\Gamma^2(-\alpha)}{\Gamma^2(\alpha+1)\Gamma^2(\alpha+2)}\,.
\ee

The approximate recursion relation Eq.~\eqref{eq:recursion} can be applied to a cascade decay with any number of final states. It  provides therefore a very simple way to estimate the rate of a cascade decay: 
A cascade decay with $n$ branchings carries a weight $a^n$.   
Importantly, $a$ turns out to be much smaller than $1$. 
Even if one takes $\lambda$ to a large value such that the exponential suppression   comes only from scalar dressing, using for example the strong coupling point Eq.~\eqref{eq:pointscal},  the $a$ coefficient still remains suppressed by two orders of magnitude. 
The calculation done here is only for a scalar interaction, however similar qualitative properties can be expected for gravitons.

Having a way to estimate cascade decay rates,  let us consider qualitatively a typical cascade decay. The propagators in the cascade decay diagram  live in region \textit{2} and are thus proportional to $p^{2\alpha}$. This  implies that large values of  $p^2$  are preferred in the kinematic integrals. This favors large energy and small 3-momenta in the outgoing legs. One thus recognizes the familiar tendency of the KK modes to decay into near-threshold states---a result expected from approximate 5d Lorentz invariance and already studied in \cite{Csaki:2008dt}. Since the boost with respect to the daughter particle is small, spatial correlations between final states are small and the cascade decays of a bulk field leads to spherical events. Still because of the $p^{2\alpha}$ factors, a decay $a\rightarrow 1 +2$ tends to prefer equal outgoing $p_{\rm daughters} \approx \frac{p}{2}$. These features provide  a qualitative picture of the most favored kinematic configuration: each KK continuum splits into two KK continuum sharing half of the absolute 4-momentum.~\footnote{This is somehow the continuous version of a KK mode with mass $m_{n}$ decaying into two modes with mass $m_{\rm daughters} \approx m_{n}/2$. }

We can use this typical kinematic configuration to get an idea of the rate as a function of the typical final state momentum. The initial field is assumed to have absolute timelike momentum $p$. 
 The decay ends at a typical absolute momentum $p_f$ which could for instance be the typical momentum for which the KK modes are long lived enough to escape the detector. 
 Assuming that the absolute momentum is divided by 2 at each branching, we have $p/p_f\sim 2^n$.
 Consider that such splittings occur $n$ times, such that the full cascade decay diagram has $2^n-1$ vertices. The square amplitude is proportional to $a^{2^n-1}$.
  As a result, the square amplitude for this cascade decay diagram is suppressed by 
\be
\Gamma\sim a^{p/p_f-1}\,. 
\ee
Since $a$ is much smaller than $1$, this continuum decay rate is  exponentially suppressed  as a function of the initial momentum. 

This last part of the analysis is of course very qualitative but it applies to the most likely kinematic configuration hence it is a good hint that  bulk cascade decays  are exponentially suppressed as a function of the momentum of the initial continuum. Numerical evaluations, perhaps using Monte Carlo integration, could be used to test this claim. 
 It would also be interesting to evaluate the ``soft bomb'' rate directly from the CFT side.

\section{Conclusion \label{se:con}}

In this note we have shown that propagators in a truncated Lorentzian AdS$_5$ background are exponentially  suppressed in the IR region of the bulk when the conformal coordinate $z$ exceeds a threshold of $O(1/|p|)$, with $p=\sqrt{p^\mu p_\mu }$. 
 While such an ``opacity'' property is trivial for spacelike momentum, for timelike momentum  we show that it is a consequence of the dressing of 5d propagators by bulk interactions.
AdS gravity induces a universal contribution to the dressing, and  we find that the leading effect comes from the scalar component of the 5d graviton. 

Exact loop calculations in AdS are notoriously difficult, and the calculations in this work  rely instead on a set of approximations,  the main ones being
\textit{(i)} the  limits of the AdS propagator in different regions of position-momentum space and
\textit{(ii)} a 5d position-momentum space  narrow-width expansion. 
   Part of the results have been checked numerically using exact expressions.  
Our results tend to be more exact at strong coupling, while for weaker coupling extra contributions to the dressing from regions \textit{3}, \textit{4}    may also matter, such that the opacity effect obtained here may be underestimated.

We find that, at least at strong coupling, the exponential suppression ``censors'' the region of position-momentum space where the 5d EFT  would become invalid. We argue that under certain conditions this effect may be valid for any  strength of 5d gravity. The properties of opacity and censorship 
discussed in this work  are inherent to curved space and do not occur in flat space.

Building on these results we turn to another feature of Lorentzian AdS: the cascade decay of bulk fields. These are  known to give soft spherical events with high multiplicity, a property
also expected from large-$N$ strongly interacting CFT sectors. 
Focusing on a cubic interaction, using the techniques previously developed and a rough approximation on threshold effects, we obtain an approximate recursion relation between cascade decay rates with $n$ and $n+1$ final states. Qualitative considerations then suggest that the rate for bulk cascade decays is  exponentially suppressed as a function of the momentum $p$ of the initial state. 
It would be interesting to pursue such study with more accurate calculations or perhaps numerical tools, both on the AdS and CFT sides. 

The set of results obtained in this note can be seen as a contribution to establishing a solid understanding  of effective field theories in AdS. 
 But one may also find it inspiring in the scope of physics beyond the Standard Model, as  it makes clear that the truncated AdS background   can be used to describe a strongly interacting dark sector which somehow vanishes at high energy. This direction will be pursued in the companion paper \cite{mypaper}.

 \section*{Acknowledgements}

I am grateful to Flip Tanedo, Csaba Cs\'aki,  David Poland, Eric Perlmutter, Mark Wise, Prashant Saraswat, Gero von Gersdorff, Eduardo Pont\'on and Philippe Brax for useful discussions and comments. I thank UCR where part of this work was realized. 
 This work is supported by the S\~ao Paulo Research Foundation (FAPESP) under grants \#2011/11973, \#2014/21477-2 and \#2018/11721-4.

\appendix

\section{Dressed propagator in KK representation \label{app:KK}}

For completeness we show how the dressed KK representation  of $\Delta_q(z,z')$ follows from the general dressed equation of motion Eq.~\eqref{eq:EOM_dressed}. This is somewhat standard but will make clear that the KK modes get generally mixed by their self-energies.

Let us introduce the matrix notation 
\be
{\bf f}(z)=\begin{pmatrix}f_n(z)\end{pmatrix}\,\quad {\bf D}=\begin{pmatrix}\frac{ \delta_{np}}{q^2-m_n^2}\end{pmatrix}
\ee
where ${\bf f}$ is a one-dimensional infinite vector and ${\bf D}$ is an infinite diagonal matrix.  

The free propagator reads
\be
\Delta^{(0)}_q(z,z')=i\,{\bf f}(z)\cdot {\bf D} \cdot  {\bf f}(z')\,.
\ee
Starting from the geometric series representation of the solution to Eq.~\eqref{eq:EOM_dressed}, 
\begin{align}
\Delta_p(z,z')=&\Delta^{(0)}_q(z,z') +  \int du \int dv \,\Delta^{(0)}_q(z,u) i\Pi(u,v) \Delta^{(0)}_q(v,z')+\ldots \\
=&\Delta^{(0)}_q(z,z') +  \int du \int dv \,  {\bf f}(z)\cdot i{\bf D}  \cdot {\bf f}(u) \, i\Pi(u,v)  \, {\bf f}(v)\cdot i{\bf D} \cdot  {\bf f}(z') +\ldots
\end{align}
we see that the matrix
\be
i {\bf \Pi}\equiv  \int du \int dv \, i\Pi(u,v) {\bf f}(u) \otimes {\bf f}(v)= \begin{pmatrix}
 \int du \int dv \, i\Pi(u,v) f_n(u) f_m(v)
\end{pmatrix}
 \,
\ee
appears. This is the self-energy matrix for the entire set of KK modes and it is in general non-diagonal. 
Keeping only the imaginary part of the self-energy and summing the series gives the dressed KK representation
\be
\Delta_q(z,z')=i\,{\bf f}(z) \cdot \left[ {\bf D}^{-1} + i{\rm Im}\,\Pi
\right]^{-1}\cdot  {\bf f}(z') \,
\ee
where $[\,\,]^{-1}$ is the matrix inverse. This makes clear that the full non-diagonal self-energy matrix has to be in principle included in the propagator, which non-trivially mixes the KK modes.

\section{Graviton loop} \label{app:grav_loop}

Let us introduce the notations
\begin{align}
\Delta^h_p(z,z') &= \Delta_p(z,z') \big|_{\alpha=2,b_i=0}\,, \quad \\
\Delta^B_p(z,z') &= \Delta_p(z,z') \big|_{\alpha=1,b_i=\infty}\,, \quad \\
\Delta^\phi_p(z,z') &= \Delta_p \big|_{\alpha=0,b_i=0}\,,\quad \\
\Delta^\chi_p(z,z') &=  -\Delta_p \big|_{\alpha=2,b_i=0}\,.
 \end{align}
The propagators of the graviton degrees of freedom in region \textit{2} and away from the poles are given by 
\be
\Delta^{h}(z,z')= i\frac{2k}{p^2}+i\frac{2\gamma-1+2\log\left(p/2k\right) - \pi \tan\left(p/\mu+\pi/4\right) }{2k}\,,
\ee
\be
\Delta^{\chi}(z,z')= - i\frac{2k}{p^2}-i\frac{2\gamma-1+2\log\left(p/2k\right) - \pi \tan\left(p/\mu+\pi/4\right) }{2k}\,,
\ee
\be
\Delta^{\phi}(z,z')=  i\frac{2\gamma+2\log\left(p/2k\right) - \pi \tan\left(p/\mu+\pi/4\right) }{2k}
\ee   
 \be
 \Delta^{B}(z,z')=i\left(\frac{1}{2k}-\frac{kz_<^2}{2}\right)
 -i\frac{p^2((kz_<)^2-1)}{8k^3}\left(
 2\gamma-1+2\log\left(p/2k\right) - \pi \tan\left(p/\mu+\pi/4\right)
 \right)\,.
 \ee

The contour integrals in region \textit{2} over a circle of radius $1/z_>$  are given by
\be
\frac{1}{-2\pi}\int_{{\cal C}[z_>^{-1}]} d\rho \, \Delta^h(\sqrt{\rho};z,z') =
\frac{1}{2k \, z_>^2}\,,
 \ee     
\be
\frac{1}{-2\pi}\int_{{\cal C}[z_>^{-1}]} d\rho \, \Delta^\chi(\sqrt{\rho};z,z')= -
\frac{1}{2k\, z_>^2 }\,,
 \ee           
 \be
\frac{1}{-2\pi}\int_{{\cal C}[z_>^{-1}]} d\rho \, \Delta^\phi(\sqrt{\rho};z,z')= 
\frac{1}{2k\, z_>^2}\,,
 \ee           
  \be
\frac{1}{-2\pi}\int_{{\cal C}[z_>^{-1}]} d\rho \, \Delta^B(\sqrt{\rho};z,z')= 
\frac{1-(kz_<)^2}{16(kz_>)^3 z_> }\,.
 \ee           
The integral of the bulk scalar away from the UV brane is given by 
\begin{align}
\frac{1}{-2\pi}\int_{{\cal C}[z_>^{-1}]} d\rho \, \Delta^\Phi(\sqrt{\rho};z,z')= 
  \frac{2k^3 z z'}{\Gamma(\beta+1)\Gamma(\beta+2)}
\left( \frac{z_<}{4 z_>}\right)^{\beta+1} \,.
\end{align}

When reducing the diagrams,  the following integrals are needed,
      \be
  {\rm Im} \left[  \frac{1}{i} \int \frac{d^4q}{(2\pi)^2}  \frac{1}{q^2(p+q)^2}  \right]    =  \frac{1}{16\pi} \,,
      \ee
\be
  {\rm Im} \left[  \frac{1}{i} \int \frac{d^4q}{(2\pi)^2}  \frac{p.(p+q)}{q^2(p+q)^2}  \right]   = -\frac{p^2}{32\pi} \,,
      \ee      
\be
  {\rm Im} \left[  \frac{1}{i} \int \frac{d^4q}{(2\pi)^2}  \frac{(p.(p+q))^2}{q^2(p+q)^2}  \right]   = \frac{p^4}{64\pi} \,.
      \ee        

The amplitudes read
\begin{align}
{\rm Im}\,& \Pi^{h+\chi}(z,z,') = \frac{1}{2\,M^3} \frac{1}{(kz)^3(kz')^3} \frac{1}{16\pi} \times \\ & \nonumber \left[\frac{p^2}{2}
  \left(
\frac{2 m^2}{(kz)^2}+\frac{2 m^2}{(kz')^2} +2 (\overleftarrow{\partial_z} \overrightarrow{\partial_z}+\overleftarrow{\partial_{z'}} \overrightarrow{\partial_{z'}}) \right) +
 \left(
\frac{2 m^2}{(kz)^2} +2 \overleftarrow{\partial_z} \overrightarrow{\partial_z} \right) 
\left(
\frac{2 m^2}{(kz')^2} +2 \overleftarrow{\partial_{z'}} \overrightarrow{\partial_{z'}} \right) \right] \\ & \nonumber \quad\quad\quad
\times\frac{1}{4\pi^2}\int_{\cal C}d\rho\, \Delta^\Phi(\sqrt{\rho},z,z') \int_{\tilde{\cal  C}}d\tilde \rho\, \Delta^\chi(\sqrt{\tilde \rho},z,z') \,
\end{align}
\begin{align}
\nonumber {\rm Im}\, \Pi^{B}(z,z,')&=\frac{1}{M^3} \frac{1}{(kz)^2(kz')^2} \frac{1}{16\pi} \left(
 p^2 \overrightarrow{\partial_z} \overrightarrow{\partial_{z'}} 
-p^2 (\overleftarrow{\partial_z} \overrightarrow{\partial_{z'}}
+
\overrightarrow{\partial_z} \overleftarrow{\partial_{z'}}
 )/2
\right) \\ & 
\times \frac{1}{4\pi^2}
\int_{\cal C}d\rho\, \Delta^\Phi(\sqrt{\rho},z,z') \int_{\tilde{\cal  C}}d\tilde \rho\, \Delta^B(\sqrt{\tilde \rho},z,z')\,,
\end{align}
\begin{align}
\nonumber {\rm Im}\, \Pi^{\phi}(z,z,')= & \frac{4}{3\,M^3} \frac{1}{(kz)(kz')}\frac{1}{16\pi} \left(
\frac{ m^2}{2(kz)^2} +\frac{3}{2} \overleftarrow{\partial_z} \overrightarrow{\partial_z} \right) 
\left(
\frac{ m^2}{2(kz')^2} +\frac{3}{2} \overleftarrow{\partial_{z'}} \overrightarrow{\partial_{z'}} \right) \\& \nonumber
\times\frac{1}{4\pi^2}\int_{\cal C}d\rho\, \Delta^\Phi(\sqrt{\rho},z,z') \int_{\tilde{\cal  C}}d\tilde \rho\, \Delta^\phi(\sqrt{\tilde \rho},z,z') \,.
\end{align}
The $\overrightarrow{\partial}$ derivatives act on the internal $\Delta^\Phi(\sqrt{\rho},z,z')$ only. The $\overleftarrow{\partial}$ would act on the external $\Phi$ propagator only. However we can formally integrate by part, replacing thus the $\overleftarrow{\partial}$ derivatives by derivatives acting on the whole loop, namely on the internal $\Delta^\Phi(\sqrt{\rho},z,z')$, $\Delta^{h,B,\phi,\chi}(\sqrt{\rho},z,z')$, and the $z$-dependent factors from the vertices. This leads to Eqs.~\eqref{eq:ImPih}-\eqref{eq:ImPiphi} where boundary terms have been neglected.

\bibliographystyle{JHEP}

\bibliography{biblio}

\end{document}